\documentclass[useAMS, usenatbib]{mn2e}

\input psfig.sty

\newcommand {\apj} {ApJ}
\newcommand {\apjl} {ApJL}
\newcommand {\apjs} {ApJS}
\newcommand {\mnras} {MNRAS}

\newcommand {\aap} {A\&A}
\newcommand {\aj} {AJ}

\newcommand {\nat} {Nature}

\newcommand {\pasp} {PASP}

\newcommand {\etal} {et~al.~}
\def \spose#1{\hbox  to 0pt{#1\hss}}  
\newcommand {\lta} {\mathrel{\spose{\lower 3pt\hbox{$\sim$}}\raise  2.0pt\hbox{$<$}}}
\newcommand {\gta} {\mathrel{\spose{\lower  3pt\hbox{$\sim$}}\raise 2.0pt\hbox{$>$}}}
\def \ion#1#2{#1{\footnotesize{#2}}\relax}
\newcommand {\ha}  {\ifmmode H\alpha \else H$\alpha $ \fi} 
\newcommand {\hi} {\ion{H}{I} } 

\newcommand {\kms} {\ifmmode  \,\rm km\,s^{-1} \else $\,\rm km\,s^{-1}  $ \fi }
\newcommand {\kpc} {\ifmmode  {\rm kpc}  \else ${\rm  kpc}$ \fi  }  
\newcommand {\pc} {\ifmmode  {\rm pc}  \else ${\rm pc}$ \fi  }  
\newcommand {\Msun} {\ifmmode M_{\odot} \else $M_{\odot}$ \fi} 
\newcommand {\yr} {\ifmmode yr^{-1} \else $yr^{-1}$ \fi} 
\newcommand {\hMsun} {\ifmmode h^{-1}\,\rm M_{\odot} \else $h^{-1}\,\rm M_{\odot}$ \fi}

\newcommand {\LCDM} {\ifmmode \Lambda{\rm CDM} \else $\Lambda{\rm CDM}$ \fi}

\newcommand {\Rvir} {\ifmmode R_{\rm vir} \else $R_{\rm vir}$ \fi}
\newcommand {\Vvir} {\ifmmode V_{\rm  vir} \else  $V_{\rm vir}$  \fi} 
\newcommand {\Mvir} {\ifmmode M_{\rm  vir} \else $M_{\rm  vir}$ \fi}  
\newcommand {\Jvir} {\ifmmode J_{\rm vir} \else $J_{\rm vir}$ \fi} 
\newcommand {\Evir} {\ifmmode E_{\rm vir} \else $E_{\rm vir}$ \fi} 
\newcommand {\lamgal} {\ifmmode \lambda_{\rm gal}  \else $\lambda_{\rm gal}$ \fi} 
\newcommand {\lamstar} {\ifmmode \lambda_{\rm star}  \else $\lambda_{\rm star}$ \fi} 
\newcommand {\lam} {\ifmmode \lambda  \else $\lambda$ \fi} 

\newcommand {\mgal}    {\ifmmode m_{\rm gal}    \else $m_{\rm gal}$ \fi} 
\newcommand {\Mstar}    {\ifmmode M_{\rm star}    \else $M_{\rm star}$ \fi} 
\newcommand {\Mgas}    {\ifmmode M_{\rm gas}    \else $M_{\rm gas}$ \fi} 
\newcommand {\Mgal}  {\ifmmode M_{\rm gal}  \else $M_{\rm gal}$ \fi}
\newcommand {\Matom}  {\ifmmode M_{\rm atom}  \else $M_{\rm atom}$ \fi}
\newcommand {\Mmole}  {\ifmmode M_{\rm mol}  \else $M_{\rm mol}$ \fi}
\newcommand {\jgal} {\ifmmode j_{\rm gal} \else $j_{\rm gal}$ \fi}  
\newcommand {\Jgal} {\ifmmode J_{\rm gal} \else $J_{\rm gal}$ \fi}  
\newcommand {\Vrot} {\ifmmode  V_{\rm rot} \else $V_{\rm rot}$ \fi} 

\newcommand {\mdotcool} {\ifmmode  \dot{M}_{\rm cool} \else $\dot{M}_{\rm cool}$ \fi} 
\newcommand {\mdotbar} {\ifmmode  \dot{M}_{\rm bar} \else $\dot{M}_{\rm bar}$ \fi}

\title[Origin of the Galaxy SFR Sequence]
{On the Origin of the Galaxy Star-Formation-Rate Sequence: Evolution and Scatter}
    
\author[Dutton \etal]
{Aaron A. Dutton$^{1,2}$\thanks{dutton@uvic.ca}\thanks{CITA National Fellow}, Frank C. van den Bosch$^3$, \& Avishai Dekel$^4$\\
  $^1$UCO/Lick Observatory, University of California, Santa Cruz, CA
  95064, USA\\ $^2$Dept. of Physics and Astronomy, University of
  Victoria, Victoria, BC, V8P 5C2, Canada\\
 $^3$Dept. of Physics and Astronomy, University of Utah, Salt Lake City, UT 84112-0830, USA\\
 $^4$Racah Institute of Physics, The Hebrew University, Jerusalem 91904, Israel\\}

\begin{document}
             
\date{submitted to MNRAS}
             
\pagerange{\pageref{firstpage}--\pageref{lastpage}}\pubyear{2009}

\maketitle           

\label{firstpage}


\begin{abstract}
  We use a semi-analytic model for disk galaxies to explore the origin
  of the time evolution and small scatter of the galaxy SFR sequence
  --- the tight correlation between star-formation rate (SFR) and
  stellar mass ($\Mstar$).
  The steep decline of SFR from $z \sim 2$ to the present, at fixed
  $\Mstar$, is a consequence of the following: First, disk galaxies
  are in a steady state with the SFR following the net (i.e., inflow
  minus outflow) gas accretion rate.  The evolution of the SFR
  sequence is determined by evolution in the cosmological specific
  accretion rates, $\propto (1+z)^{2.25}$, but is found to be
  independent of feedback. Although feedback determines the outflow
  rates, it shifts galaxies along the SFR sequence, leaving its zero
  point invariant.
  Second, the conversion of accretion rate to SFR is materialized
  through gas density, not gas mass. Although the model SFR is an
  increasing function of both gas mass fraction and gas density, only
  the gas densities are predicted to evolve significantly with redshift.
  Third, star formation is fueled by molecular gas. Since the
  molecular gas fraction increases monotonically with increasing gas
  density, the model predicts strong evolution in the molecular gas
  fractions, increasing by an order of magnitude from $z=0$ to $z\sim
  2$. On the other hand, the model predicts that the effective surface
  density of atomic gas is $\sim 10 \Msun \pc^{-2}$, independent of
  redshift, stellar mass or feedback.
  Our model suggests that the scatter in the SFR sequence reflects
  variations in the gas accretion history, and thus is insensitive to
  stellar mass, redshift or feedback.  The large scatter in halo spin
  contributes negligibly, because it scatters galaxies along the SFR
  sequence. An observational consequence of this is that the scatter
  in the SFR sequence is independent of the size (both stellar and
  gaseous) of galaxy disks.
\end{abstract}


\setcounter{footnote}{1}


\section{Introduction}
\label{sec:intro}

Understanding the star formation history of the universe is a major
goal of modern astronomy.  Starting at redshift $z=0$, the comoving
cosmic star formation rate (SFR) density (in units of $\rm \Msun \,
yr^{-1} \,Mpc^{-1}$) rises by an order of magnitude to $z\simeq 1$
(e.g. Lilly \etal 1996; Schiminovich \etal 2005; Hopkins \& Beacom
2006; Villar \etal 2008, Sobral \etal 2009), reaches a peak around
$z\simeq 2$ and declines to higher redshifts (Madau \etal 1998;
P{\'e}rez-Gonz{\'a}lez \etal 2008; Bouwens \etal 2008).

A number of studies have shown that the decline in the comoving cosmic
SFR density from $z\simeq 1$ to the present is not due to a decline in
the major merger rate (Bell \etal 2005; Wolf \etal 2005; Melbourne
\etal 2005; Lotz \etal 2008; Sobral \etal 2009). Rather, since the
majority of all stars seem to form in galactic disks (at least below
$z\simeq 2$), it is generally believed that the drop in comoving
cosmic SFR density from $z\simeq 1$ to the present is related to
'quiescent' processes, e.g., gas consumption and/or a reduced rate of
accretion of fresh gas from the cosmic web.

\begin{figure*}
\centerline{
\psfig{figure=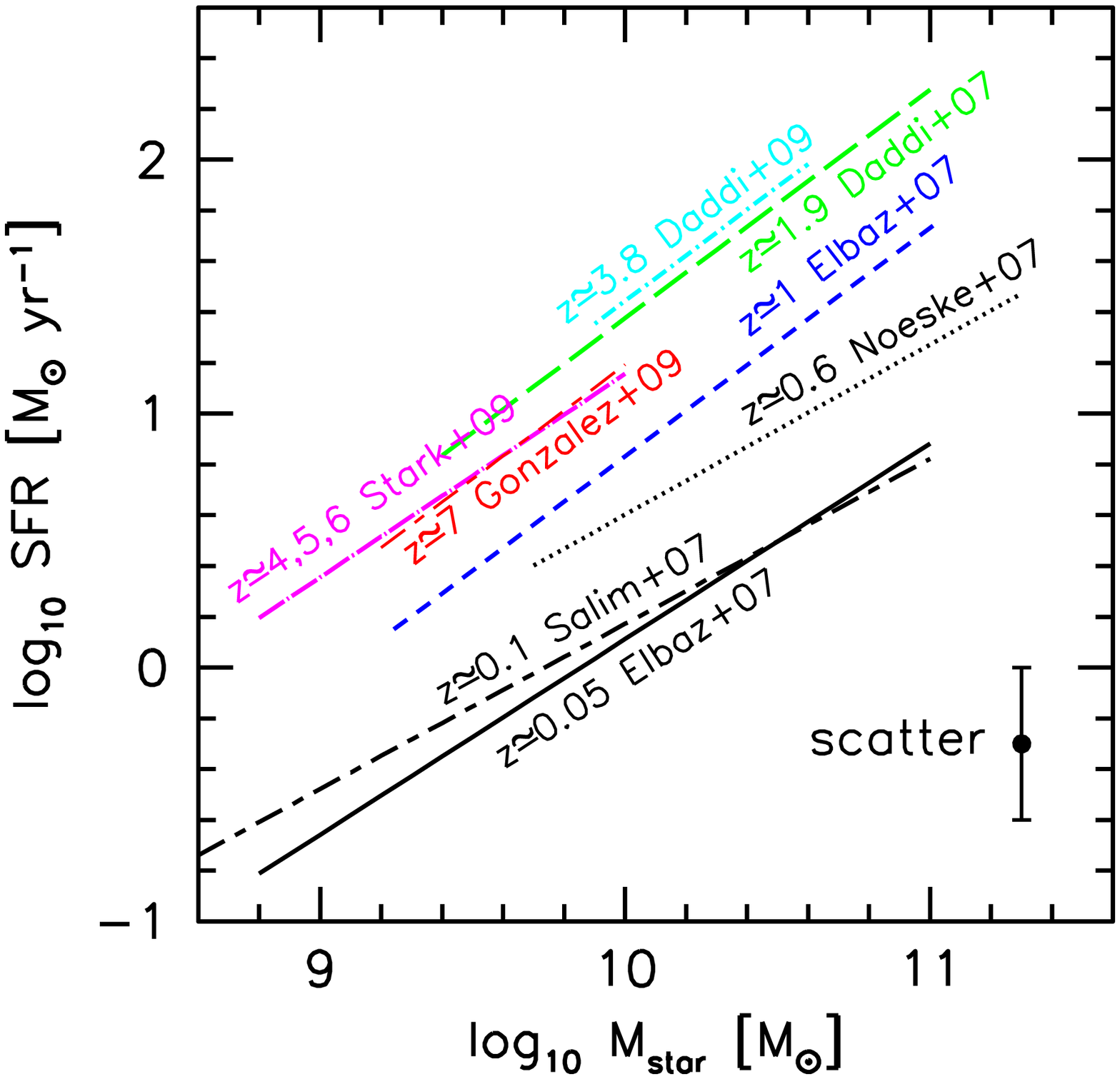,width=0.49\textwidth}
\psfig{figure=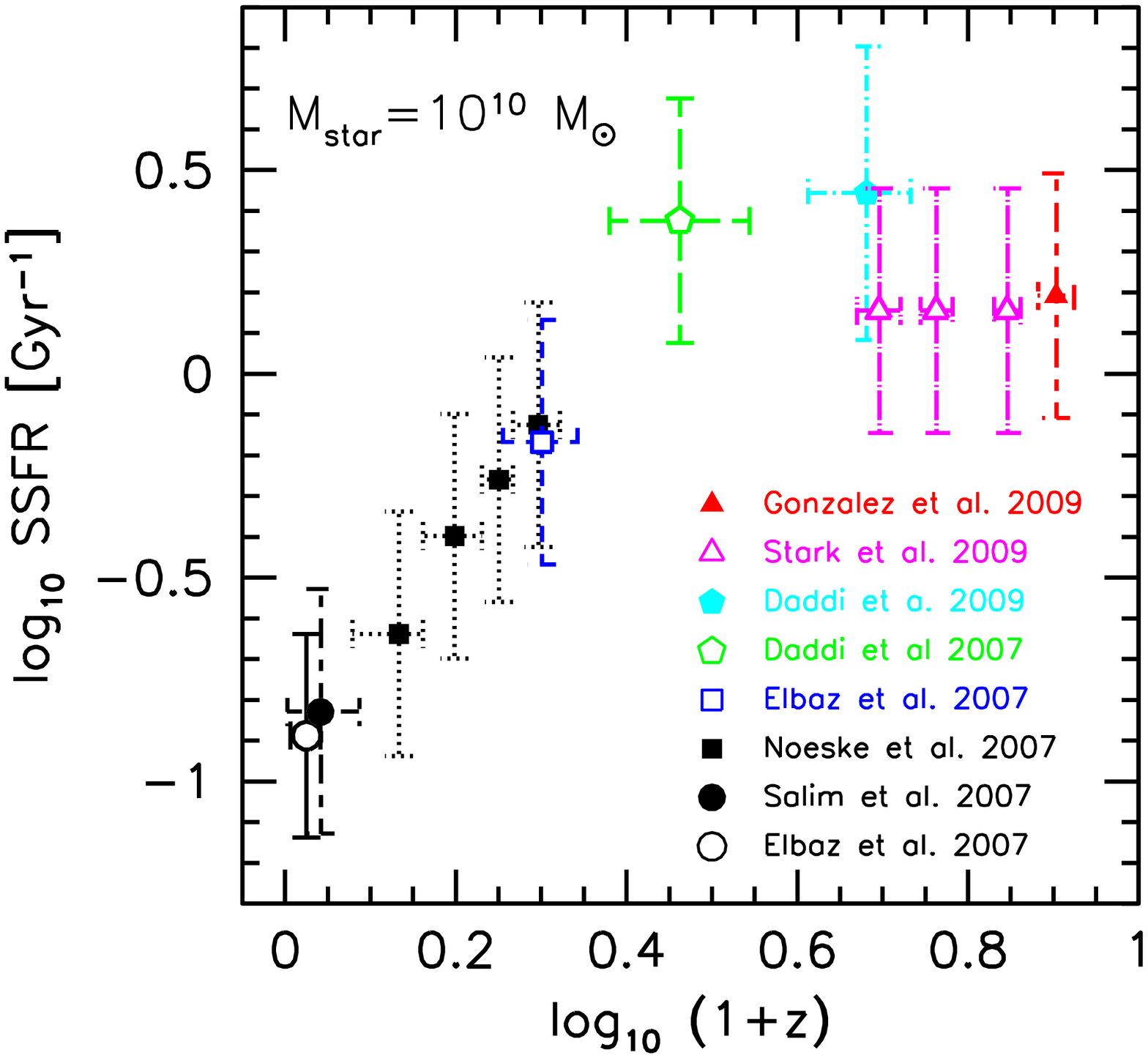,width=0.49\textwidth}}
\caption{Summary of observations from redshifts $z\simeq 0$ to
  $z\simeq 7$. Left-hand panel: SFR sequence at different
  redshifts. Right-hand panel: redshift evolution of the specific star
  formation rate (SSFR=SFR/$\Mstar$) at a stellar mass of
  $\Mstar=10^{10}\Msun$. The horizontal error bars indicate the
  redshift range of the data points, while the vertical error bars
  indicate the estimated intrinsic scatter in SSFR.  }
\label{fig:obs}
\end{figure*}

Great progress is currently being made in measuring molecular gas
masses in high redshift galaxies (e.g. Tacconi \etal 2006, 2008, 2010;
Daddi \etal 2008, 2009b). These studies indicate that the ratio
between molecular gas and stellar mass was higher in the
past. However, uncertainty in the conversion factor between CO
luminosity and molecular gas mass limits the current accuracy of these
measurements. In addition there are at present no direct observations
of atomic gas masses or gas accretion rates in high redshift galaxies,
and thus the origin of the decline in SFRs since $z\simeq 2$ is not
yet understood.

A step forward in the characterization of the evolution of galaxy SFRs
was the discovery of a relatively tight correlation between galaxy SFR
and stellar mass ($\Mstar$), with a slope just below unity and an
intrinsic scatter $\lta 0.3$ dex (Noeske \etal 2007a; Elbaz \etal
2007; Daddi \etal 2007; Salim \etal 2007).  Throughout this
paper we refer to this correlation as the SFR sequence.  The zero
point appears roughly constant from $z\sim 7$ to $z\simeq 2$ (Daddi
\etal 2009a; Stark \etal 2009; Gonzalez \etal 2009), and evolves to
lower SFR from $z\simeq 2$ to $z=0$ (Daddi \etal 2007).  See
Fig.~\ref{fig:obs} and Table~\ref{tab:obs} for a summary of the
observational data. The scatter of the SFR sequence appears to be
  invariant with redshift, while there is some indication that the
  slope is shallower at lower redshifts.

The two main questions that we address in this paper are:
\begin{enumerate}
\item What is the physical cause of the decline in the specific star
formation rate (SSFR=SFR/$\Mstar$) since $z\simeq 2$?
\item Why does the SFR sequence have such a small scatter?
\end{enumerate}

A plausible explanation for the decline in SSFR since $z\simeq 2$ is a
decline in the gas accretion rate onto galaxies. For dark matter
haloes the specific accretion rate scales as $(1+z)^{2.25}$ for $z
\lta 2$ (Birnboim, Dekel, \& Neistein 2007). Furthermore, scatter in
the gas accretion rates from cosmological simulations is about 0.3 dex
(Dekel \etal 2009). Thus the decline in SSFR as well as the small
scatter could be driven entirely by the properties of gas accretion.
However, for this to be the case, a mechanism is needed to explain why
the star formation rate should follow the gas accretion rate.

To get some insight into this problem, we assume that the star
formation rate surface density follows a Schmidt-Kennicutt (SK,
Schmidt 1959, Kennicutt 1998) law:
\begin{equation}
\Sigma_{\rm SFR} = \epsilon_{\rm SF} \Sigma_{\rm gas}^n,
\end{equation}
and that the gas surface density follows an
exponential distribution: 
\begin{equation}
\Sigma_{\rm gas}(R) = \Sigma_{0} \exp(-R/R_{\rm gas}), 
\end{equation}
where $\Sigma_0$ is the central surface density of the gas, and
$R_{\rm gas}$ is the scale length of the gas disk. The SFR is then
given by
\begin{eqnarray}
SFR & = & 2 \pi \int_{0}^{\infty} \Sigma_{\rm SFR}(R) R\, dR\\
                 & = & \epsilon_{\rm SF} M_{\rm gas}\, n^{-2}\, \Sigma_0^{n-1}\; (n>0)
\end{eqnarray}
Thus, for $n>1$, the mechanism which converts higher accretion rates
into higher star formation rates is either an increase in the gas
masses or an increase in the gas surface densities.

There is a significant scatter ($\simeq 0.14-0.2$ dex) in the sizes of
galactic stellar disks at a given stellar mass or luminosity
(e.g. Shen \etal 2003; Courteau \etal 2007), which corresponds to a
scatter in disk surface density of $\simeq 0.28-0.4$ dex. If the
surface density of gas is correlated with that in stars, then the
observed amount of scatter in the surface densities of stellar
galactic disks would imply a scatter of 0.14-0.2 dex in SSFR (assuming
$n=1.5$), and even larger scatter for a higher $n$. Thus galaxy size
could be an important contributor to the scatter in the SFR sequence.

\begin{table*}
  \caption{Summary of the observed SFR sequence from low to high redshift. 
    The relation is of the form SFR = SFR$_0 (\Mstar/10^{10}\Msun)^{\rm
      slope}$. All stellar masses and SFRs have been converted to a Chabrier IMF. Parameters measured by ourselves are indicated by a *.}
  \begin{tabular}{llllllll}
    \hline 
    \hline  
    Redshift & Redshift range &  SFR$_0$ & slope & Mass range & original IMF & Reference\\
   & &  $[\rm M_{\odot}\, yr^{-1}]$ & & $\log_{10}$ $[M_{\odot}]$ & & \\
    \hline
    0.06 & 0.015-0.10 & 1.29   & 0.77 & 8.8-11.1 & Salpeter & Elbaz \etal (2007) \\
    0.10 & 0.005-0.22 & 1.48   & 0.65 & 8.0-11.3 & Chabrier & Salim \etal (2007) \\
    0.36 & 0.20-0.45 & $2.3^*$ & 0.67 & 9.7-11.3 & Chabrier & Noeske \etal (2007a) \\
    0.58 & 0.45-0.70 & $4.0^*$ & 0.67 & 9.7-11.3 & Chabrier & Noeske \etal (2007a) \\
    0.78 & 0.70-0.85 & $5.5^*$ & 0.67 & 10.0-11.3 & Chabrier & Noeske \etal (2007a) \\
    0.98 & 0.85-1.10 & $7.5^*$ & 0.67 & 10.0-11.3 & Chabrier & Noeske \etal (2007a) \\
    1.0  & 0.80-1.20 & 6.80     & 0.90 & 9.3-11.1 & Salpeter & Elbaz \etal (2007) \\
    1.9  & 1.4-2.5 & 23.77      & 0.90 & 9.5-11.1 & Salpeter & Daddi \etal (2007) \\
    3.8  & 3.1-4.4 & 29.3      & 0.90 & 9.9-10.6 & Chabrier & Daddi \etal (2009a) \\
    3.96 & 3.67-4.25 & $14.3^*$  & $0.8^*$ & 8.8-10.3 & Salpeter & Stark \etal (2009) \\
    4.79 & 4.54-5.04 & $14.3^*$  & $0.8^*$ & 8.8-10.1 & Salpeter & Stark \etal (2009) \\
    6.01 & 5.76-6.26 & $14.3^*$  & $0.8^*$ & 9.1-9.8 & Salpeter & Stark \etal (2009) \\
    7.3  & 6.9-7.7 & 15.52      & $0.9^*$ & 9.2-10.0 & Salpeter & Gonzalez \etal (2009) \\
    \hline 
    \hline
\label{tab:obs}
\end{tabular}
\end{table*}

In this paper we use the semi-analytic disk galaxy evolution model of
Dutton \& van den Bosch (2009) to study the evolution and scatter of
the SFR sequence. In \S 2 we give a brief overview of the galaxy
evolution model.  In \S 3 we discuss the evolution of the zero point
of the SFR sequence, and whether it is driven by gas accretion rate,
gas fractions or gas density. The scatter in the SFR sequence is
addressed in \S 4, while \S 5 discusses why the SFR sequence is
independent of feedback. We summarize our results in \S 6.


\section{Disk Galaxy Evolution Model}
\label{sec:models}

Here we give a brief overview of the disk galaxy evolution model used
in this paper. A more detailed description of this model is given in
Dutton \& van den Bosch (2009; hereafter DB09).

The key difference with traditional semi-analytic models
(e.g. Kauffmann \etal 1993; Cole \etal 1994,2000; Somerville \&
Primack 1999; Hatton \etal 2003; Cattaneo \etal 2006; Croton \etal
2006) is that in this model the inflow (due to gas cooling), outflow
(due to supernova driven winds), star formation rates, and
metallicities, are all computed {\it as a function of radius}, rather
than being treated as global parameters. This allows us to follow self
consistently the growth of gaseous and stellar galactic disks, and to
split the gas disk into atomic and molecular phases. Our model should
thus be considered in the context of previous models of this type
(e.g. Kauffmann 1996; Avila-Reese \etal 1998; Firmani \& Avila-Reese
2000; van den Bosch 2001,2002).

The main assumptions that characterize the framework of our models are
the following:
\begin{enumerate}

\item  Dark matter  haloes around  disk  galaxies grow  by the  smooth
  accretion of mass which we model with the Wechsler \etal (2002) mass
  accretion history  (MAH). The shape of  this MAH is  specified by the
  concentration of the halo at redshift zero;

\item Dark matter haloes are assumed to be spherical and to follow a
  density distribution given by the NFW profile (Navarro, Frenk, \&
  White 1997), which is specified by two parameters: mass and
  concentration. However, halo concentration is known to be correlated
  with halo mass, and we use mass dependence and evolution of the
  concentration parameter given by the Bullock \etal (2001a) model
  with parameters (Macci\`o \etal 2008) for a WMAP 5th year cosmology
  (Dunkley \etal 2009);

\item Gas that enters the halo is shock heated to the virial
  temperature, and acquires the same distribution of specific angular
  momentum as the dark matter (van den Bosch \etal 2002).  We use the
  angular momentum distributions of the halo as parametrized by
  Bullock \etal (2001b) and Sharma \& Steinmetz (2005);

\item Gas cools radiatively, conserving its specific angular momentum,
  and forms a disk in centrifugal equilibrium. Cooling rates are
  computed using the cooling functions of Sutherland \& Dopita (1993)
  under the assumption of collisional ionization equilibrium;

\item Star formation is assumed to take place in dense molecular gas.
  In order to compute the density of molecular gas, we follow Blitz \&
  Rosolowsky (2006) and relate the molecular gas fraction to the
  midplane pressure in the disk. For completeness, a detailed
  description of our star formation prescription is given in
  Appendix~A, where we show that it yields a SK star formation law
  that is truncated at low $\Sigma_{\rm gas}$, in good agreement with
  observations;

\item Supernova (SN) feedback re-heats some of the cold gas, ejecting
  it from the disk and halo. We assume that ejected gas is lost
    forever.  We consider two feedback models, one driven by energy
    (e.g. Dekel \& Silk 1986), the other by momentum (e.g. Murray,
    Quataert, \& Thompson 2005). For both models we assume that the outflow
    moves at the local escape velocity of the disk-halo system,
    $V_{\rm esc}(R)$. The key difference between these two models is
    the dependence of the mass loading factor, $\eta$ (ratio between
    outflow mass rate and star formation rate), on wind velocity. For
    the energy driven wind $\eta\propto V_{\rm esc}^{-2}$, whereas for
    the momentum driven wind $\eta \propto V_{\rm esc}^{-1}$. Note
    that since the escape velocity typically varies with
    galacto-centric radius, so do does the mass loading factor.

\item Stars eject metals into  the inter-stellar medium, enriching the
  cold gas;

\item Bruzual  \& Charlot  (2003) stellar population  synthesis models
  are convolved with the star formation histories and metallicities to
  derive photometric fluxes in different pass bands.
\end{enumerate}
Each model galaxy  consists of five mass components:  dark matter, hot
halo gas, disk mass in stars,  disk mass in cold gas, and ejected gas.
The  dark matter  and the  hot gas  are assumed  to be  distributed in
spherical shells,  the disk mass  is assumed to be  in infinitesimally
thin annuli.  Throughout this paper we  refer to $R$ as radius, $t$ as
time (where $t=0$ is defined as the Big Bang) and $z$ as redshift.

For each galaxy we set up a radial grid with 200 bins quadratically
sampled from between 0.001 and 1 times the redshift zero virial
radius.  As described in DB09 for the purpose of conserving angular
momentum it is more convenient to use a grid in specific angular
momentum, $j$.  We convert the initial grid in $R$, to a grid in $j$
using $j^2/G = R M(R)$, where $M(R)$ is the halo mass enclosed within
a sphere of radius $R$.  We follow the formation and evolution of each
galaxy using 400 redshift steps, quadratically sampled from $z=10$ to
$z=0$.  For each time step we compute the changes in the various mass
components in each radial bin.

\subsection{Limitations of the Model}
Our model makes several simplifying assumptions about the physical
processes that occur during disk galaxy formation (mergers, gas
accretion, star formation, feedback, adiabatic contraction). In some
cases it is because the correct physics is unknown, in other cases it
is because modeling all of the details adds extra complication to the
model without necessarily improving our understanding.  A discussion
of these assumptions and the limitations they imply to our model are
given in DB09. Here, and in Appendix B, we focus on the issue of gas
accretion, and specifically the differences between the classical
``hot mode'' vs. the new ``cold mode'' accretion models.

Our assumption about the way gas is accreted into galaxies (by a
cooling flow of shock heated has) is likely incorrect. Simulations
suggest that disk galaxies accrete most of their mass through cold
flows, and that in the absence of extra heating or outflows the baryon
fraction of galaxies is close to the universal value (Birnboim \&
Dekel 2003; Keres \etal 2005; Dekel \& Birnboim 2006). As we
discuss in Appendix B, for haloes with masses below $\sim
10^{12}\Msun$, the cooling time is short compared to the free fall
time, and thus the accretion rates onto the central galaxy are the
same in the classical cooling model as in the cold mode model.  The
key difference between these two accretion models occurs for high mass
$\Mvir \gta 10^{12} \Msun$ haloes at high $z\gta 2$ redshifts. In this
regime the accretion rates are low in the classical cooling model
because cooling is inefficient, but the accretion rates can be high in
the cold mode model because accretion rates are dominated by cold
streams penetrating the hot halo (Dekel \etal 2009).  Thus our results
for the cold gas accretion rates onto galaxies with stellar
masses less than about $5\times 10^{10}\Msun$ are not expected to be
effected by us ignoring the phenomenon of cold streams.

It is possible that the angular momentum acquired by galaxies may be
different in the hot mode and cold mode accretion models. However,
non-radiative cosmological hydrodynamical simulations (e.g. van den
Bosch \etal 2002; Sharma \& Steinmetz 2005) have shown that the
specific angular momentum of the gas and dark matter halo are very
similar, which can be understood from the fact that both have
experienced the same torques. Thus as long as most of the baryons
accrete onto the central galaxy (as is the case for galaxies in haloes
below $\sim 10^{12}\Msun$) then the available angular momentum is
expected to be the same in the hot mode and cold mode accretion
models.

\begin{figure*}
\centerline{
\psfig{figure=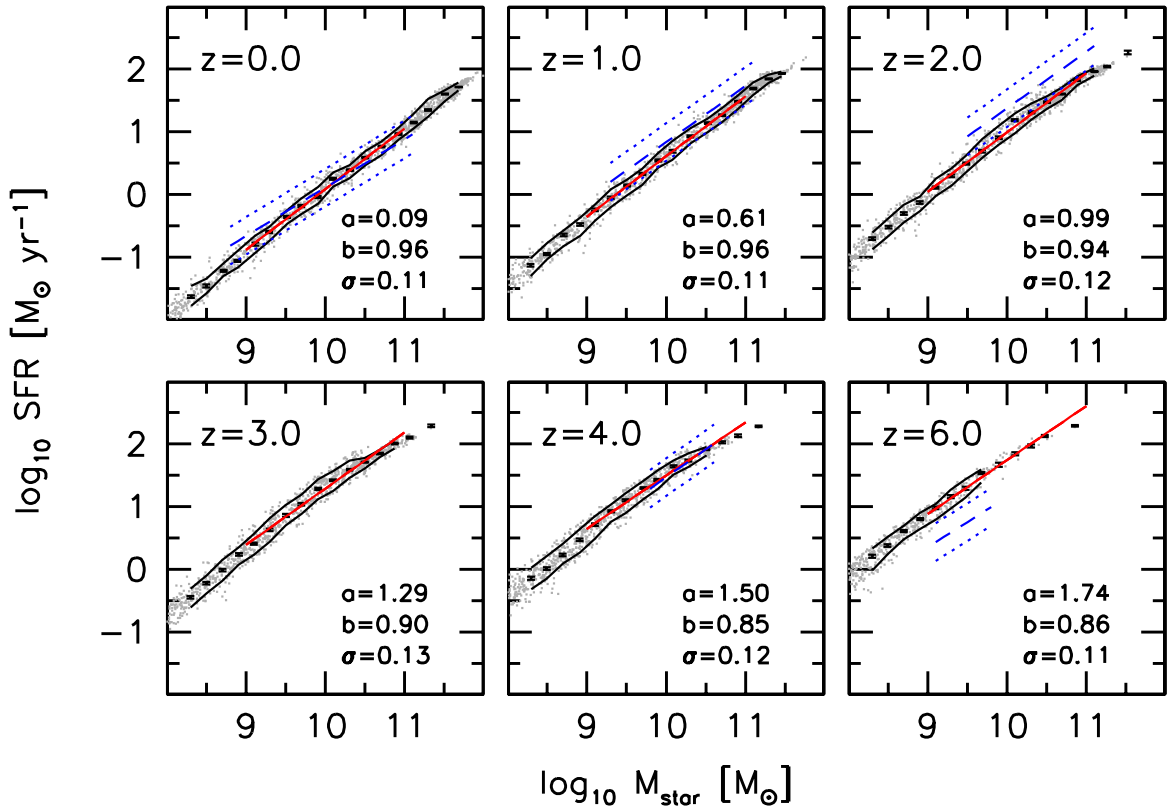,width=0.90\textwidth}}
\caption{The SFR sequence for our EFB model with energy driven
  outflows. The model galaxies are represented by grey points, the
  black error bars show the error on the median SFR in $\Mstar$ bins
  of 0.2 dex width, and the black lines enclose 68.3\% of the model
  galaxies. The blue dashed lines show the observed relations (only
  plotted over the mass range for which observations exist), with the
  dotted blue lines showing the observed scatter of 0.3 dex.  The red
  lines show a fit to the model galaxies in the range $9 \leq
  \log_{10} M_{\rm star}/[M_{\odot}] \leq 11.0$. The parameters of
  these fits are given in the lower right corner of each panel. The
  slope and scatter of the model SFR sequence is roughly independent
  of redshift, whereas the zero point evolves strongly.}
\label{fig:sfrm-efb}
\end{figure*}

\subsection{Overview of model parameters}

In this paper we consider three models that only differ in their
treatment of feedback and the amount of angular momentum loss. In
addition to a model without feedback (NFB), we consider a model with
energy driven feedback (EFB) and one with momentum driven feedback
(MFB). The parameters of these models have been tuned as little as
possible.  The NFB model has no tuning. The MFB and EFB models have
their respective feedback efficiencies and amount of angular momentum
loss tuned to reproduce the zero points of the rotation velocity -
stellar mass and stellar disk size - stellar mass relations at
redshift zero.  For the MFB model this results in maximal feedback and
no angular momentum loss, for the EFB model this results in 25\%
feedback efficiency and 30\% angular momentum loss. Details of these
models, as well as the resulting galaxy scaling relations (rotation
velocity - stellar mass, stellar disk size - stellar mass, gas
fraction - stellar mass, gas metallicity - stellar mass) and disk
structural properties, are discussed in DB09 and Dutton (2009). As a
summary, the MFB and EFB models reproduce the scaling relations
reasonable well, while the NFB model does not reproduce any of the
scaling relations.

The input parameters of our models are as follows.

(1) Cosmology: $\Omega_{\rm m,0}, \Omega_{\Lambda}$, $\Omega_{\rm b}$,
  $\sigma_8$, $h$, $n$.  In this paper we adopt a flat \LCDM cosmology
  motivated by the 5th year WMAP results (Dunkley \etal 2009). In
  particular, we adopt $\Omega_{\rm m,0}=0.258,
  \Omega_{\Lambda}=0.742, \Omega_{\rm b}=0.044, \sigma_8=0.80$,
  $h=0.7$, and $n=1$.
    
(2) Dark Halo structure: $K, F, \sigma_{\log c}$. We assume that the
  concentrations of dark matter haloes of a given mass, $M$, follow a
  log-normal distribution with median $c(M)$ from the Bullock \etal
  (2001a) model, parametrized by $F=0.01$ and $K=3.7$, and scatter
  $\sigma_{\log c}=0.11$. Whenever this model yields $c < K$ we set
  $c=K$. This parametrization nicely reproduces the distribution of
  halo concentrations of relaxed dark matter haloes in cosmological
  N-body simulations (e.g. Wechsler \etal 2002; Macci\`o \etal 2007,
  2008; Zhao \etal 2009).

  (3) Angular momentum distribution: $\bar{\lambda}$, $\sigma_{\ln
    \lambda}$, $\alpha$, $\sigma_{\log\alpha}$. We assume that the
  spin parameters of dark matter haloes of a given mass, $M$, follow a
  log-normal distribution with median $\bar{\lambda}$ and scatter
  $\sigma_{\ln \lambda}=0.5$ (Bullock \etal 2001b). For our models
  with no outflows and momentum driven outflows we adopt the
  cosmological value for relaxed haloes $\bar{\lambda}=0.035$
  (Macci\`o \etal 2008), while for our model with energy driven
  outflows we assume that 30\% of the angular momentum has been lost,
  so that $\bar{\lambda}=0.025$. This choice of $\bar{\lambda}$ is
  made to reproduce the zero point of the stellar disk size - stellar mass
  relation (see DB09), and is motivated by the fact that baryons tend
  to lose specific angular momentum in hydrodynamical cosmological
  simulations (e.g. Navarro \& Steinmetz 2000; Piontek \& Steinmetz
  2009). We assume that the specific angular momentum distribution of
  both dark matter and gas in a halo of given mass and spin parameter
  is given by the fitting function presented in Sharma \& Steinmetz
  (2005), which is parametrized by a single parameter, $\alpha$.
  Based on their numerical simulations, we assume that the
  distribution of $\alpha$ is log-normal, with a median $\alpha=0.9$
  and scatter $\sigma_{\log\alpha}=0.11$. We note that the
  parametrization of the angular momentum distribution by Bullock
  \etal (2001b), which is specified by $\mu$, results in very similar
  disk density profiles as the Sharma \& Steinmetz (2005)
  parametrization, except for the extremes of the distribution of
  $\alpha$ and $\mu$ (see Dutton 2009).

(4) Star formation: $\tilde \epsilon_{\rm SF}$. Our star formation
  prescription (see Appendix~A), is parametrized by a single free
  parameter, $\tilde \epsilon_{\rm SF}$, which sets the normalization
  of $\Sigma_{\rm SFR}$ at a given surface density of dense molecular
  gas. We tune this parameter so that we reproduce the normalization of
  the empirical SK star formation law, which gives
  $\tilde\epsilon_{\rm SF}=13 \rm \,Gyr^{-1}$.

  (5) Feedback: $\epsilon_{\rm EFB}$, $\epsilon_{\rm MFB}$, $\eta_{\rm
    SN}$, $E_{\rm SN}$, $p_{\rm SN}$. For the energy driven wind model
  the mass loading factor is given by $\eta= 2\epsilon_{\rm EFB}\,
  E_{\rm SN}\,\eta_{\rm SN}/V_{\rm esc}^2$, while for the momentum
  driven wind model $\eta= \epsilon_{\rm MFB}\,p_{\rm SN}\,\eta_{\rm
    SN}/V_{\rm esc}$. We adopt an energy per supernova of $E_{\rm
    SN}=10^{51}$ erg, a momentum per supernova of $p_{\rm SN}=3\times
  10^{4} \Msun\kms$, and a supernova rate of $\eta_{\rm
    SN}=8.3\times10^{-3}$ per Solar mass of stars formed.  For our
  energy driven outflow model we adopt $\epsilon_{\rm EFB}=0.25$
  (i.e. 25\% of the SN energy ends up in the outflow) and for our
  momentum driven outflow model we adopt $\epsilon_{\rm MFB}=1$
  (i.e. 100\% of the SN momentum ends up in the outflow). The no
  feedback model has $\epsilon_{\rm EFB}= \epsilon_{\rm MFB}=0$.

\begin{figure*}
\centerline{
\psfig{figure=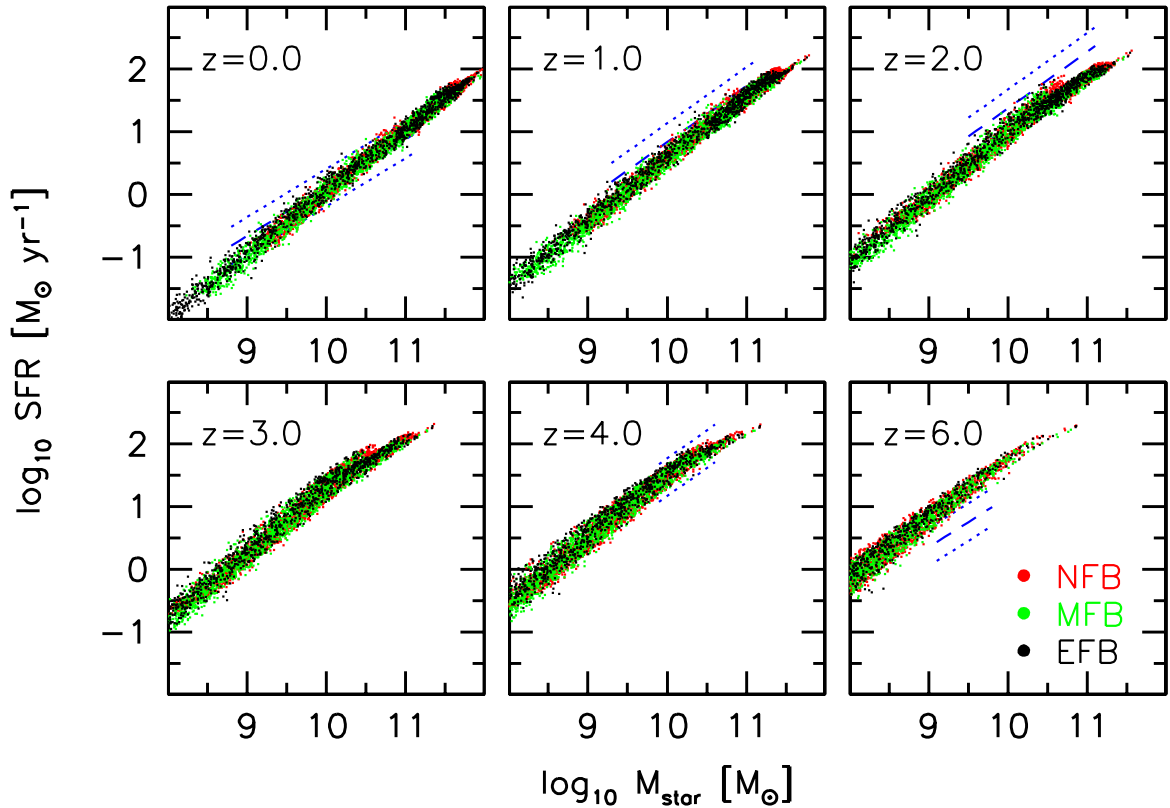,width=0.90\textwidth}}
\caption{The SFR sequence for different feedback models: no feedback
  (red dots), momentum driven feedback (green dots) and energy driven
  (black dots). The slope, zero point and scatter of the SFR sequence
  is independent of feedback (the different colored dots trace the
  same relations).}
\label{fig:sfrmstar}
\end{figure*}
    
(6) Stellar populations and chemical enrichment: ${\cal R}, y, Z_{\rm
  hot}$, and the choice of initial mass function (IMF).  We assume
  that the baryonic gas that is accreted by a dark matter halo has a
  Helium mass abundance of 0.25 and an initial metallicity of $Z_{\rm
  hot} = 0.002 \simeq 0.1 Z_{\odot}$. Contrary to DB09, where we
  adopted the instantaneous recycling approximation (IRA), here we
  calculate the amount of recycled gas as a function of time by
  convolving the star formation history with the return fractions from
  the BC03 models. At low redshifts this makes no difference to the
  final stellar masses, but at high redshifts the IRA over-estimates
  the return fraction and hence under-estimates the stellar masses.
  For simplicity, we still use the IRA for the {\it chemical}
  evolution, using a stellar yield of $y=0.02$ appropriate for the
  Chabrier (2003) IMF adopted throughout.

\section{Origin of the Zero-Point Evolution of the SFR Sequence}
\label{sec:evol}

\subsection{Summary of observations}
\label{sec:obs}

The observational measurements of the SFR sequence from low to high
redshifts that we use in this paper are summarized in Table
\ref{tab:obs} and Fig.~\ref{fig:obs}. The slope of the SFR sequence
defined as ${\rm d}\log SFR/{\rm d}\log \Mstar$ is $ \simeq
0.65-0.95$. Shallower slopes tend to be measured at low redshifts
(Elbaz \etal 2007; Noeske \etal 2007; Salim \etal 2007; Schiminovich
\etal 2007; Bothwell \etal 2009). Studies above $z\simeq 1$ tend to
find steeper slopes of around 0.9, in particular the recent study by
Pannella \etal (2009) finds a slope of $0.95\pm0.07$ at $z\simeq 2$
for stellar masses in the range $\Mstar \simeq
10^{10.2-11.0}\Msun$. It is not clear if this implies an evolution of
the slope, or is due to a systematic difference in SFR indicators or
sample selection used at different redshifts.  The intrinsic scatter
is somewhat uncertain, but is estimated to be $\lta 0.3$ dex
(e.g. Noeske \etal 2007a; Elbaz \etal 2007; Daddi \etal 2007).  The
zero-point evolves by a factor of $\sim 20$ from redshift $z\simeq 2$
to $z\simeq 0$, but is roughly constant from $z\sim 7$ to $z\simeq
2$. Note, though, that the high redshift $(z \gta 4)$ data have not
been corrected for extinction as observations indicate that there is
little or no extinction in these galaxies (Gonzalez \etal
2009). Nevertheless, we caution that the observed SFRs at $z\gta 4$
may have been underestimated.

These relations assumed either a Salpeter IMF or a Chabrier (2003)
IMF, as indicated in Table~\ref{tab:obs}. For consistency, we correct
the stellar masses and SFRs to a Chabrier (2003) IMF by subtracting
0.25 dex (e.g. Gonzalez \etal 2009).  Given that the slope of the SFR
sequence is close to unity, both Salpeter and Chabrier (2003) IMF's
result in very similar zero-points to the SFR sequence.

\begin{figure*}
\centerline{
\psfig{figure=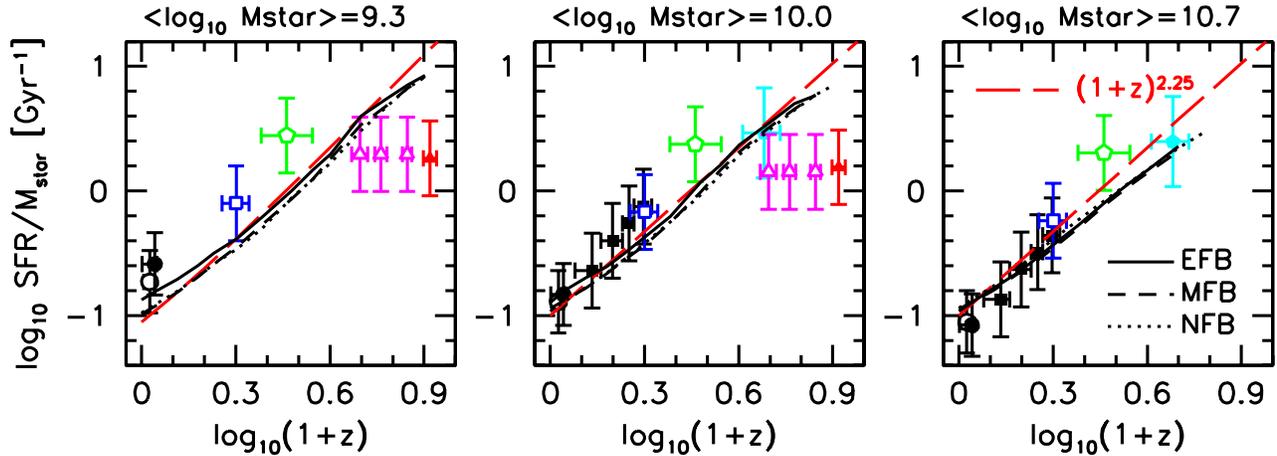,width=0.99\textwidth}}
\caption{ Evolution in the specific star formation rates for stellar
  masses of $\log_{10} \Mstar /[M_{\odot}] =$ 9.3 (left panel), 10.0
  (middle panel), and 10.7 (right panel). The observed evolution is
  indicated by symbols (see Fig.~\ref{fig:obs} for references). The
  horizontal error bars indicate the redshift range of the data
  points, the vertical error bars indicate the estimated intrinsic
  scatter in SSFR.  Black lines correspond to our models: energy
  driven outflows (EFB, solid lines), momentum driven outflows (MFB,
  dashed lines), and no outflows (NFB, dotted lines). The evolution in
  the models is the same for all three feedback models, and closely
  follows the $(1+z)^{2.25}$ scaling (red long-dashed lines) for the
  dark matter accretion rates at fixed halo mass expected in a \LCDM
  cosmology. The models are in reasonable agreement with the data,
  although they seem to under predict the SSFRs at $z \simeq 2$, and
  over predict those at $z \gta 4$.}
\label{fig:sfrmz}
\end{figure*}

\subsection{Comparison with observations} 

{Fig.~\ref{fig:sfrm-efb}} shows the SFR sequence at redshifts
$z=0,1,2,3,4$ and $6$ for our model with energy driven feedback (EFB).
The red solid lines show power-law fits to the SFR sequence for
stellar masses between $10^9 \le \Mstar \le 10^{11} \Msun$, with slope
($b$), zero point ($a$) and scatter ($\sigma$) as indicated. The black
error bars show the median SFR in stellar mass bins of width 0.2
dex. The solid black lines enclose 68\% of the model data in each
stellar mass bin, and indicate that the scatter in independent of
mass. The model galaxies have been uniformly sampled in redshift zero
halo mass (with $10^{10} \le \Mvir \le 10^{13.35} h^{-1}\Msun$). At
high redshifts the tail of galaxies to high stellar masses is caused
by scatter in the MAH. In particular, the highest mass galaxies at
high redshifts ($z\gta 2$) are all in early forming (high
concentration) haloes.

The observed relations are shown by blue lines. Dashed lines show the
mean, dotted lines show the $1\sigma$ scatter, where we have adopted a
scatter of 0.3 dex.  Note that the model scatter is significantly
smaller than this ($0.12\pm 0.01$ dex), but is independent of mass and
redshift. We do not consider this a serious problem for the model,
given that the systematic uncertainties in the data are likely to be
significant (i.e., the observed scatter has to be considered an upper
limit on the intrinsic scatter).  In addition, it is likely that our
model underestimates the true scatter due to our simplified treatment
of the halo mass accretion history (i.e. the assumption that there is
a one-to-one relation between MAH and halo concentration). For
example cosmological hydrodynamical simulations find a scatter of
about 0.3 dex in the gas accretion rates in haloes of mass
$10^{12}\Msun$ at $z=2.5$ (Dekel \etal 2009). Such a scatter in gas
accretion rates would explain the full observed scatter in the SFR
sequence.

The slope of the SFR sequence in our model is close to but less than
unity, and evolves to slightly lower values at higher redshifts. In
fact, the model slope seems to be somewhat too steep compared to the
data at $z=0$. It remains to be seen whether this discrepancy holds up
with better data coming available. 

{Fig.~\ref{fig:sfrmstar}} shows the SFR sequence for our three
feedback models: no feedback (red); momentum driven feedback (green);
and energy driven feedback (black).  This shows that the slope, zero
point and scatter of the SFR sequence are all independent of the
feedback model! This feature is remarkable because every other galaxy
scaling relation depends quite significantly on the detailed treatment
of feedback (e.g. van den Bosch 2002; DB09; Dutton 2009).  The origin
of this intriguing characteristic of the SFR sequence is discussed in
\S\ref{sec:fb} below.

{Fig.~\ref{fig:sfrmz}} shows the evolution in the specific star
formation rate (SFR/$M_{\rm star}$) at three stellar masses:
$\Mstar=2\times 10^9 \Msun$ (left-hand panel), $\Mstar=1\times
10^{10}\Msun$ (middle panel), and $\Mstar=5\times 10^{10}\Msun$
(right-hand panel).  The data (described above) are shown as symbols,
while the black lines indicate our models (solid line, EFB; dashed
line, MFB; dotted line, NFB). Note that the models predict substantial
evolution in the SSFRs, in rough overall agreement with the data, and
that there is very little difference between the three different
models (as also seen in Fig.~\ref{fig:sfrmstar}). There is some
indication that the models systematically under predict the SSFRs at
$1 \lta z \lta 2$, especially for the less massive galaxies. This may
signal a potential shortcoming of the models, though it seems also
possible that the observations have over-estimated the SSFRs due to
sample selection effects or other systematics (see Dav\'e 2008 for a
discussion of systematic effects in SFRs and stellar masses).

\begin{figure*}
\centerline{
\psfig{figure=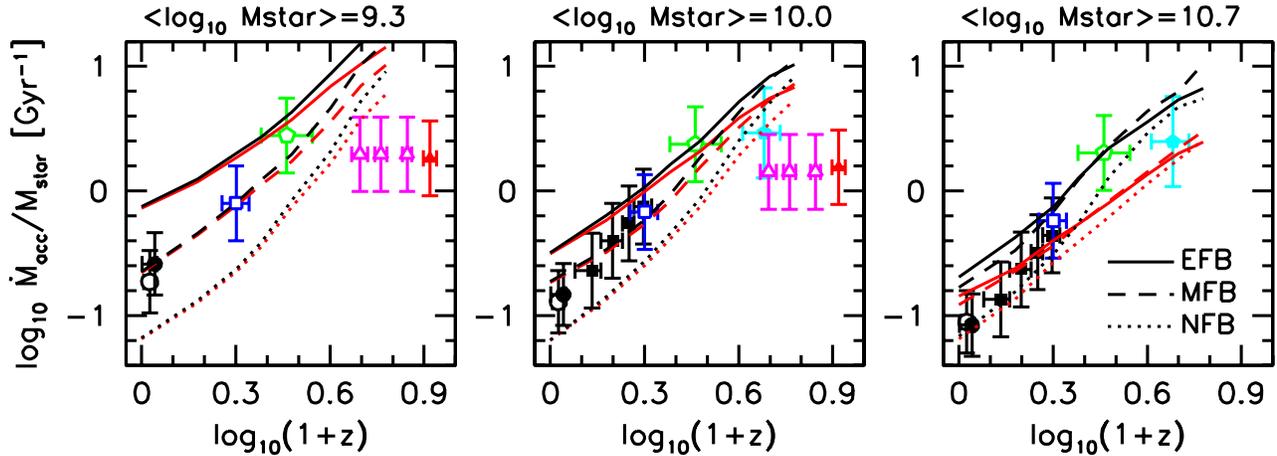,width=0.99\textwidth}}
\caption{Evolution in the specific baryon accretion rates for stellar
  masses of $\log_{10} \Mstar /[M_{\odot}] =$ 9.3 (left panel), 10.0
  (middle panel), and 10.7 (right panel). The observed evolution in
  specific star formation rates is given by the colored symbols (as in
  Fig.~\ref{fig:obs}). The specific baryon accretion rates,
  $\dot{M}_{\rm acc}/\Mstar$, into dark matter haloes are given by the
  black lines, and the specific baryon accretion rates onto the
  galaxies are given by red lines. The line types correspond to
  different feedback models: energy driven outflows (EFB, solid
  lines), momentum driven outflows (MFB, dashed lines), and no
  outflows (NFB, dotted lines). The evolution in baryon accretion
  rates closely follows the evolution in specific star formation
  rates. However, the zero point depends on the feedback model,
  especially at lower stellar masses.}
\label{fig:dmacmz}
\end{figure*}

The discrepancies between the SSFRs in our models and the data at
$z\simeq 2$ are of similar (but smaller) magnitude and direction as
found by Dav\'e (2008), using cosmological hydrodynamical simulations.
Dav\'e (2008) proposed that an evolving IMF could reconcile the data
with theory.  A key prediction of this evolving IMF model is that the
discrepancy between the observed and predicted (with fixed IMF) SSFRs
should increase monotonically with redshift. However, since the work
by Dav\'e (2008) a number of observational studies have presented
SSFRs at $z \geq 4$ (Daddi \etal 2009a; Stark \etal 2009; Gonzalez
\etal 2009).  In particular, the study of Daddi \etal (2009a) measured
the SFR sequence for B-band dropouts at $z\simeq 4$, which were found
to be only marginally higher than those measured at $z\simeq 2$. Our
models match this $z\simeq 4$ data point remarkably well.  This
suggests that the discrepancy at $z\simeq 2$ is not due to a variable
IMF, at least not as in the model advocated by Dav\'e (2008).

Current observations indicate that the SSFRs are constant from
redshifts $z\sim 7$ to $z\simeq 2$. This is on stark contrast with our
models, which predict that the SSFRs decrease monotonically with
decreasing redshift. Even though the models reproduce the observed
SSFRs at $z\simeq 4$, at $z=7$ the discrepancy between the models and
the observations is a factor of $\simeq 4$. As mentioned in
\S\ref{sec:obs} these very high-redshift data have not been corrected for
dust extinction, and thus if there is substantial extinction in very
high redshift galaxies (as there is in lower redshift galaxies), then
this could reconcile observation with theory.

\begin{figure*}
\centerline{
\psfig{figure=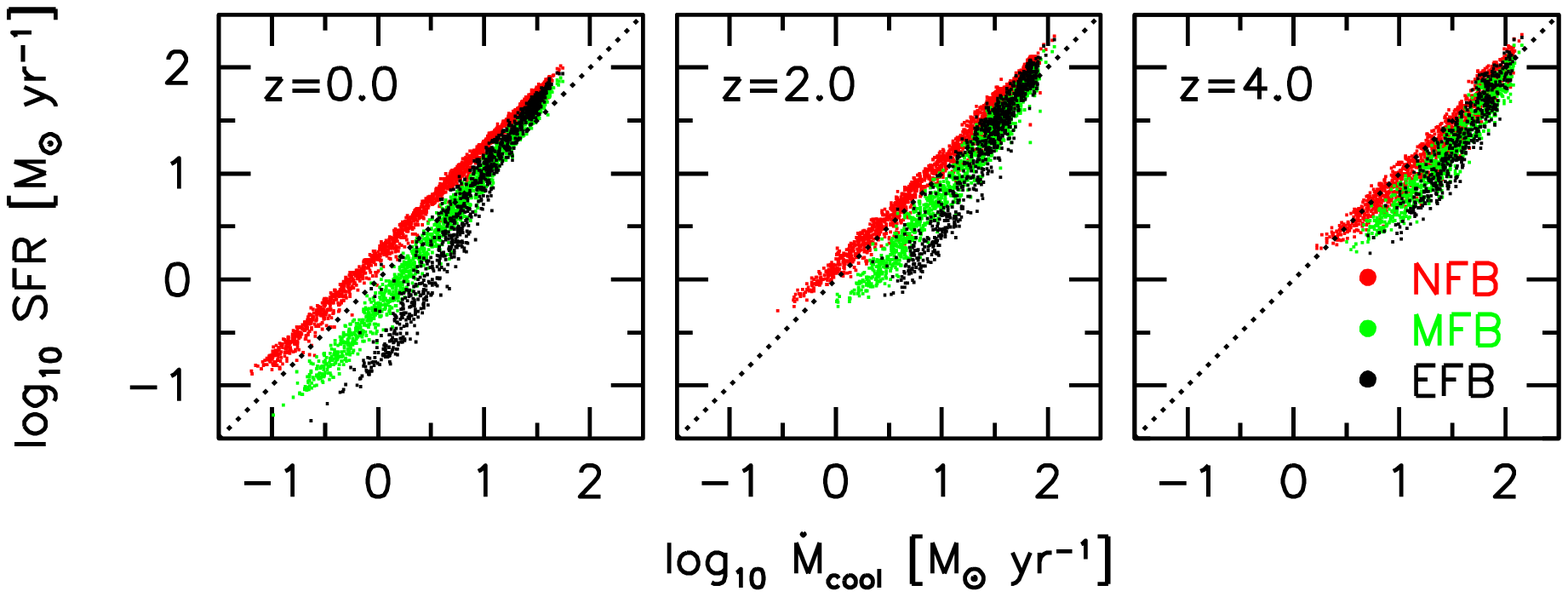,width=0.99\textwidth}}
\centerline{
\psfig{figure=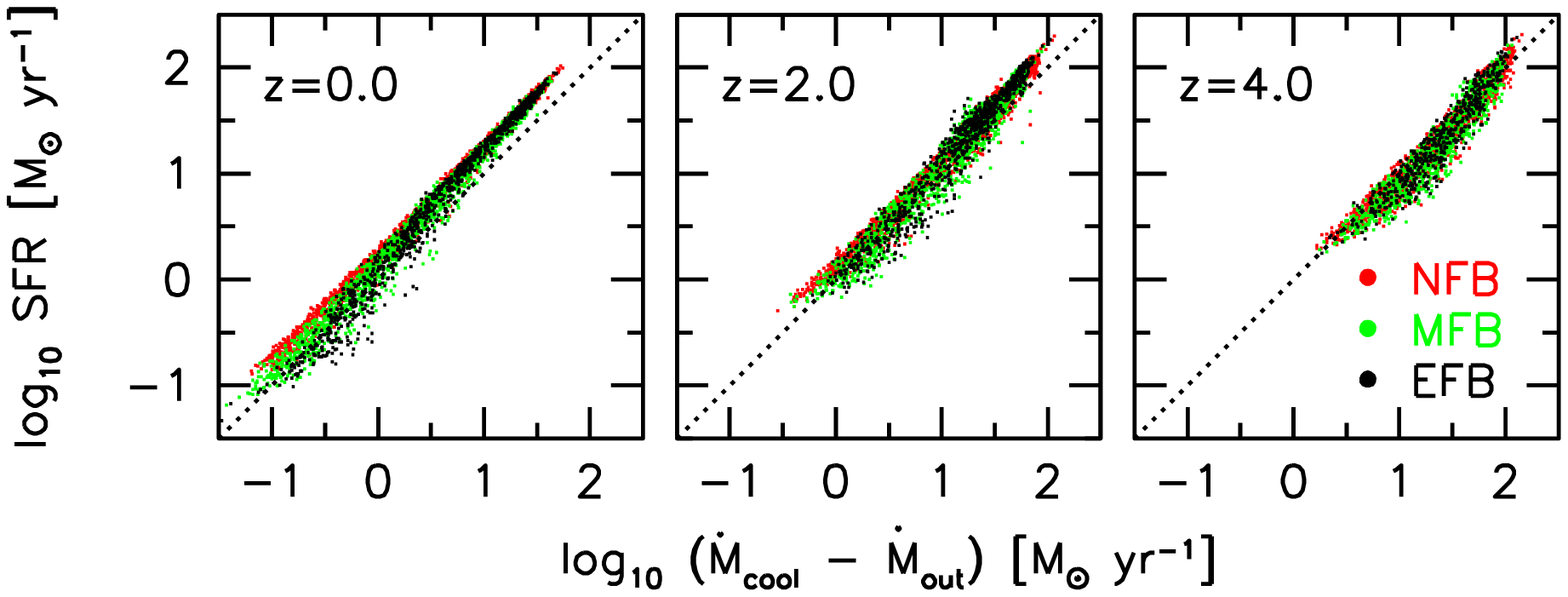,width=0.99\textwidth}}
\caption{The SFR as function of the rate at which cold gas is accreted
  by the galaxy ($\dot{M}_{\rm cool}$; upper panels) and as a function
  of the {\it net} inflow rate of cold gas onto the galaxy, given by
  $\dot{M}_{\rm cool} - \dot{M}_{\rm out}$ (lower panels). Results are
  shown for model galaxies with stellar masses $\Mstar > 10^9 \Msun$
  and for our three feedback models: energy driven outflows (EFB,
  black dots), momentum driven outflows (MFB, green dots), and no
  outflows (NFB, red dots). For the NFB model the SFR is proportional
  (but not identically equal to) the cold gas inflow rate. For models
  with outflows, the SFR is not directly proportional to the gas
  inflow rate. For these models the SFR is better correlated with the
  {\it net} gas inflow rate. This indicates that our model galaxies
  have settled in a steady state between the rate of inflowing gas and
  the rate of star formation.}
\label{fig:sfrdmcl}
\end{figure*}

\subsection{What drives the evolution in the zero point of the SFR sequence?}

Having established that our models reproduce the decrease in SSFRs by
a factor of $\sim 20$ from redshifts $z\sim 4 $ to $z=0$, we now
address the question: what drives this evolution?

\subsubsection{Gas accretion rate?}

It has been pointed out by previous authors (e.g. Noeske \etal 2007b)
that the decline of the SFRs at a given \Mstar is compatible with the
cosmological decline in the accretion rate onto dark matter haloes,
which scales roughly as $\simeq (1+z)^{2.25}$ at a fixed halo mass
(Birnboim, Dekel, \& Neistein 2007). Note that the actual slope of the
specific accretion rate varies from 2 at $z=0$ to 2.5 at high redshift
(Neistein \& Dekel 2008).

{Fig.~\ref{fig:dmacmz}} shows the evolution of the baryon accretion
rate onto the halo ($\dot{M}_{\rm acc}$) vs. redshift (black lines)
and the evolution of the cold gas accretion rate onto the galaxy
($\dot{M}_{\rm cool}$) (red lines).  As expected from {
  Figs.~\ref{fig:tcool} \& \ref{fig:mmdot}}, these two curves only
differ significantly at high stellar masses ($\Mstar \gta 5 \times
10^{10} \Msun$), where the cooling time becomes larger than the free
fall time. We note that cold streams are likely to eliminate this
difference.  For comparison, the observed evolution in star formation
rates are shown as colored symbols (as in {Fig.~\ref{fig:sfrmz})}.
The main outcome is that the evolution in the baryon/cold gas
accretion rate closely follows the evolution in the star formation
rate (both for the models as well as in the data).  This confirms that
a plausible explanation for the decline in the SSFRs of galaxies is
simply a decline in the specific gas accretion rate.  However,
the zero point is dependent on the feedback model, especially for
lower stellar masses.  This is because feedback is more efficient at
removing baryons from lower mass haloes. For a given stellar mass, the
halo mass will be higher for models with stronger feedback. Since
higher mass haloes have higher baryon accretion rates (in absolute
terms), the models with stronger feedback have higher specific baryon
accretion rates.

The upper panels of {Fig.~\ref{fig:sfrdmcl}} show the SFR vs. the cold
gas accretion rate onto the galaxy, $\dot{M}_{\rm cool}$, for our
three feedback models. For the NFB model without outflows, there is an
almost 1:1 correspondence between the SFR and inflow rate. However,
for models with outflows (MFB and EFB), the relation between SFR and
inflow rate has a slope greater than unity, with a steeper slope for
the EFB model than for the MFB model.  The differences between the
three models are largest at low accretion rates, which correspond to
low mass haloes, and hence where the specific outflow rates are
highest.  This suggests that the {\it net} inflow rate might correlate
better with the SFR. This is indeed the case, as shown in the lower
panels of {Fig.~\ref{fig:sfrdmcl}}.  Thus we conclude that galaxy
disks are in a steady state in which the star formation rate is
roughly equal to the {\it net} gas inflow rate. This conclusion can
also be derived by solving the differential equation for change in gas
mass with time (Bouche \etal 2009).

\begin{figure*}
\centerline{
\psfig{figure=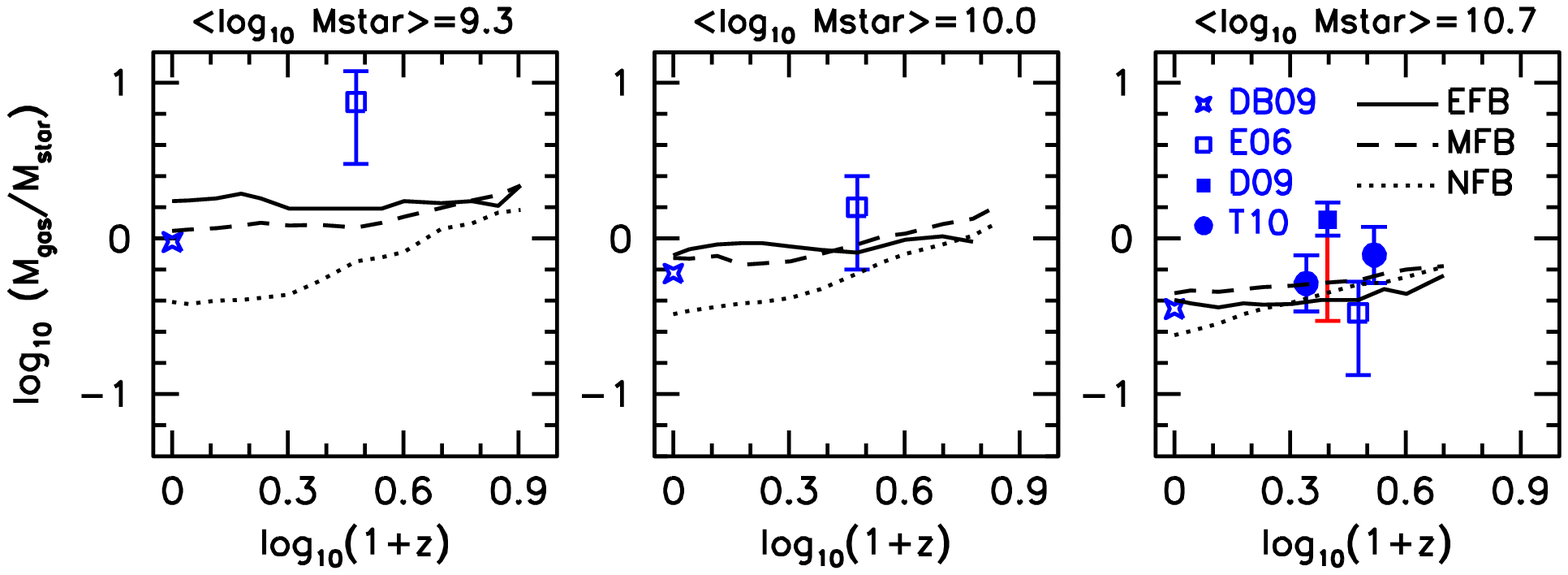,width=0.99\textwidth}}
\caption{ Evolution in the ratio between cold gas and stellar mass for
  stellar masses of $\log_{10} \Mstar /[M_{\odot}] =$ 9.3 (left
  panel), 10.0 (middle panel), and 10.7 (right panel).  The models are
  given by black lines as indicated. The observed gas-to-star ratios
  are given by blue symbols as indicated. At low redshift the data
  reflects direct measurements of atomic and molecular gas (from DB09
  using data from Garnett 2002), and are in good agreement with the
  energy and momentum driven wind models. At $z=2$ the data reflect
  gas mass fraction derived indirectly using the Schimdt-Kennicutt
  star formation law (Erb \etal 2006, E06). At $z=1.5$ the gas
  fractions are based on CO measurements and dynamical models (Daddi
  \etal 2009b, D09). The small (blue) error bar gives the nominal
  uncertainty while the large (red) error bar shows the uncertainty in
  the conversion factor from CO luminosity to molecular gas mass. At
  $z=1.2$ and $z=2.3$ the gas fractions are based on CO measurements
  (Tacconi \etal 2010, T10).  The evolution in the model gas
  fractions is weak, especially for the two models with outflows.  We
  conclude that the evolution in the SFR sequence is not dominated by
  evolution in the cold gas mass fractions.}
\label{fig:gsmxz}
\end{figure*}

\begin{figure*}
\centerline{
\psfig{figure=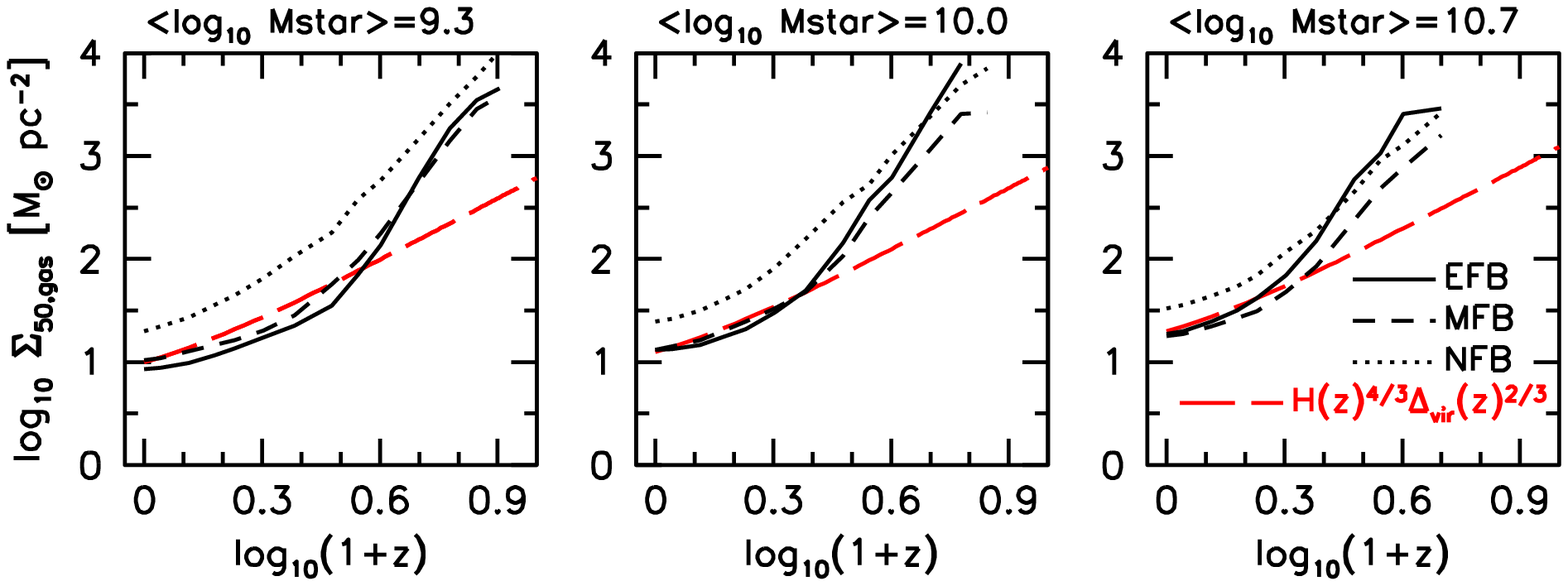,width=0.99\textwidth}}
\caption{ Evolution in the effective surface density of the cold gas
  (defined as the average density of cold gas within the cold gas half
  mass radius) for stellar masses of $\log_{10} \Mstar /[M_{\odot}] =$
  9.3 (left panel), 10.0 (middle panel), and 10.7 (right panel). The
  models are given by black lines, as indicated. There is strong
  evolution in the effective gas densities for all three models and
  for all three stellar masses. The red dashed lines show the
  evolution of the halo surface density at fixed halo mass. Given the
  weak evolution in the gas-to-stellar mass ratios
  (Fig.~\ref{fig:gsmxz}), it is this evolution in the gas surface
  densities that provides the mechanism for the evolution in gas
  accretion rates to turn into an evolution in the SFR sequence.}
\label{fig:sgasmz}
\end{figure*}

\subsubsection{Gas masses or gas densities?}

We have shown that the decline in net gas accretion rate can explain
{\it why} SSFRs decline with time, but this does not explain {\it how}
the decline in gas accretion rate is converted into a decline in the
SSFRs.  Recall that, as discussed in the introduction, the evolution
of the star formation rate at a given stellar mass is expected to
depend on the change in gas mass and/or gas density at that stellar
mass. A factor 20 decrease in SFR needs a factor 20 decrease in gas
mass, or a factor 400 decrease in gas surface density (assuming a SK
star formation law with $n=1.5$) or factor of 20 decrease in gas
  surface density (for a SK law with $n=2$). Note that while the
  global SK law has a slope of $n\simeq 1.5$, for spiral galaxies the
  spatially resolved SK law typically has $n\simeq 2$ (Bigiel \etal
  2008, and references therein) as thus it may be the more appropriate
  value to use to estimate the required evolution in gas disk
  density.

\begin{figure*}
\centerline{
\psfig{figure=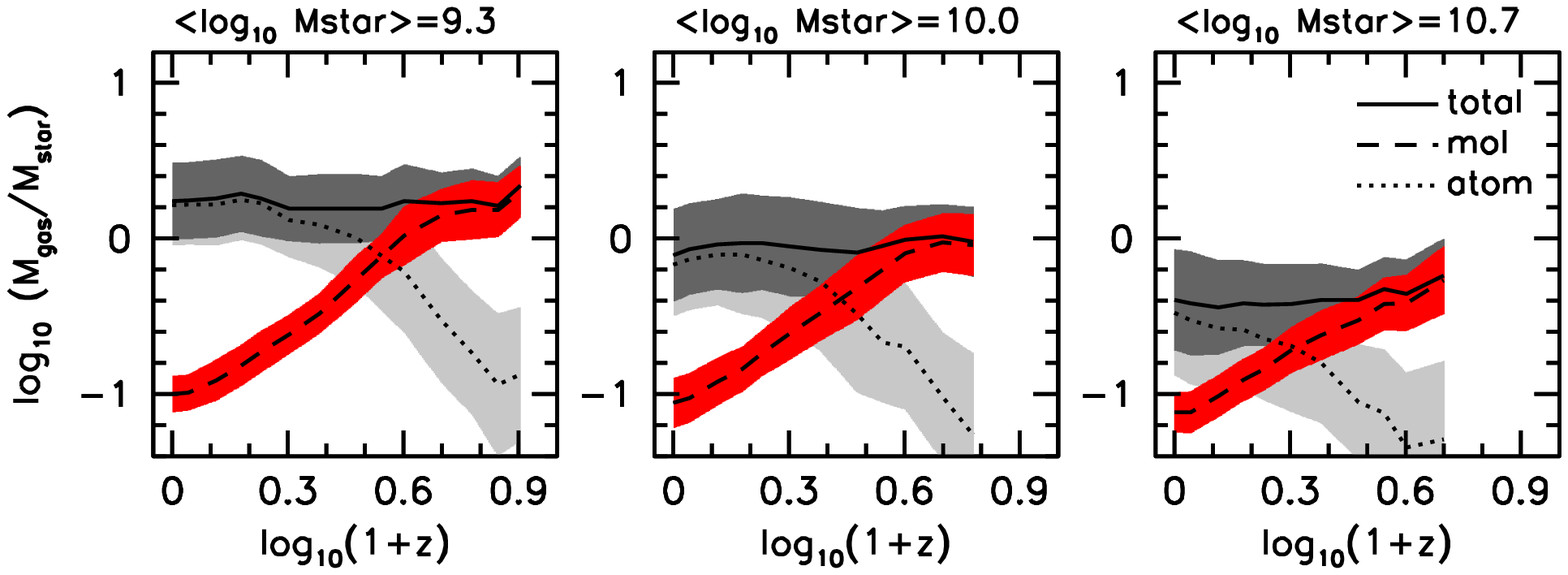,width=0.99\textwidth}}
\caption{ Evolution in the cold gas-to-stellar mass ratios for the EFB
  model, where the line types correspond to the atomic (dotted, light
  grey), molecular (dashed, red) and total (atomic plus molecular;
  solid, dark grey). The three panels correspond to stellar masses of
  $\log_{10} \Mstar /[M_{\odot}] =$ 9.3 (left panel), 10.0 (middle
  panels), and 10.7 (right panels). At fixed stellar mass, the
  molecular gas masses increase by an order of magnitude (independent
  of the details of gas accretion and feedback), while the atomic gas
  masses decrease by a similar amount. Since molecular gas is the fuel
  for star formation, and the SFR $\sim M_{\rm mol}$, the evolution in
  the molecular gas masses at fixed stellar mass translates directly
  into evolution in the specific star formation rates.}
\label{fig:gsmxz2}
\end{figure*}

{Fig.~\ref{fig:gsmxz}} shows the evolution in the cold gas to stellar
mass ratio, $M_{\rm gas}/M_{\rm star}$, where cold gas is both atomic
and molecular.  In all models the evolution is weak, especially for
the two models with outflows (solid and dashed lines). The outflow
models also reproduce the cold gas fractions at redshift $z=0$,
whereas the no feedback model (dotted lines) systematically
under-predicts the cold gas masses at all stellar masses. At high
redshifts the observations are not direct and thus carry significant
uncertainties. At $z\simeq 2$, Erb \etal (2006) derive gas masses from
the SK star formation law. In addition to the systematic uncertainties
(such as uncertainties in gas disk sizes, star formation rates, and
stellar masses) inherent in these measurements there are concerns
about sample selection. Since the Erb \etal (2006) sample is UV
selected, it may be biased towards gas rich, high star formation rate
galaxies, especially at lower stellar masses.  Given these
uncertainties, we consider our models are not necessarily in
disagreement with the observations.

At $z\simeq 1.5$, Daddi \etal (2009b) detected the CO[2-1] line for 6
star forming galaxies, and inferred a mean gas mass fraction of $57\pm
6\%$. These gas fractions are higher than predicted by our model.
However, there is a large uncertainty (a factor of $\simeq 4$) in the
conversion factor, $\alpha_{\rm CO}$, between CO luminosity and
molecular gas mass. Daddi \etal (2009b) attempt to infer $\alpha_{\rm
  CO}$ using dynamical models, but these involve many assumptions,
including the (unknown) dark matter fraction, and thus do not robustly
constrain $\alpha_{\rm CO}$. We note that even with the high
$\alpha_{\rm CO}$ advocated by Daddi \etal (2009b) this only results
in a factor of $\simeq 3$ decrease in gas masses between $z\simeq 1.5$
and $z=0$. A factor of $\simeq 10$ decrease in gas masses is required
for change in gas fractions to explain the factor of $\simeq 10$
decrease in SSFR between $z=0$ and $z\simeq 1.5$.

More recent observations (Tacconi \etal 2010) have detected
the CO[3-2] line for 19 massive ($\Mstar \simeq 5 \times 10^{10} -
3\times 10^{11} \Msun$) star forming galaxies at redshifts $z\simeq
1.2$ and $z\simeq 2.3$, finding mean molecular gas fractions of $0.34$
and $0.44$, respectively. These measurements fall between those of Erb
\etal (2006) and Daddi \etal (2009b), and are thus also consistent
with our models.

We conclude that the current observations of gas fractions are
consistent with weak or no evolution between $z\sim 2$ and $z=0$, as
our models predict.  But obviously this needs to be confirmed with
{\it direct} observations of atomic and molecular gas masses at high
redshifts, for larger samples of star forming galaxies and over a
wider range of stellar masses.

{Fig.~\ref{fig:sgasmz}} shows the evolution in the effective gas
surface density, which we define as the mean gas density within the
radius enclosing half of the cold gas mass: $\Sigma_{50,\rm gas} =
M_{\rm gas}/(2\pi R_{50,\rm gas}^2)$, where $R_{50,\rm gas}$ is the
half mass radius of the gas. In our models the gas density shows very
strong evolution: a factor of $\sim 10$ from $z=2$ to $z=0$, and a
factor of $\sim 100$ from $z=4$ to $z=0$.  It is thus the increase in
the {\it density} of the gas that provides the mechanism for turning
the higher net gas accretion rates at high redshifts into higher star
formation rates, at least in our models.

\subsubsection{Why are high redshift disks denser?}
\label{sec:dense}
Obviously this poses the question why gas disks at higher redshifts
have higher surface densities. A crude approximation to the evolution
of disk sizes, $R_{\rm d}$, can be made by assuming that 1) galaxies
have flat rotation curves and 2) that the disk spin parameter,
$\lambda_{\rm d}$, is independent of redshift. In that case, the disk
size scales with the size of the dark matter halo according to $R_{\rm
  d}(z) \sim \lambda_{\rm d} \Rvir(z)$ (Mo, Mao, \& White 1998). The
size of the halo scales with mass and redshift as $\Rvir(z)\sim
H(z)^{-2/3}\Delta_{\rm vir}(z)^{-1/3} \Mvir(z)^{1/3}$. Here $H(z)$ is
the Hubble parameter and $\Delta_{\rm vir}(z)$ is the overdensity of
the halo with respect to the critical density of the Universe.  Thus
{\it if} the size and mass of the gas disk are proportional to those
of the halo, then the density of the gas disk scales as $\Sigma(z)\sim
H(z)^{4/3}\Delta_{\rm vir}(z)^{2/3}$ at fixed halo mass. This scaling
is shown (with arbitrary normalization) as the red long-dashed lines
in {Fig.~\ref{fig:sgasmz}}. Our model roughly follows this scaling at
$z \lta 1-2.5$ (depending on stellar mass). For higher redshifts,
however, our model predicts an evolution in density that is
significantly stronger. This indicates that there are other effects in
our model that drive the evolution of gas densities besides simply
halo density evolution. As discussed in Appendix~\ref{sec:C} the
predominant additional effect is a decrease in the disk specific
angular momentum at high redshifts.

It should be noted that in our models the gas densities are higher at
higher redshift because the half mass radii are smaller. An equivalent
effect could in principle be achieved if gas disks at high redshift
were considerably more clumpy than low redshift gas disks of the
same stellar mass. The clumpiness could result in a larger fraction of
the gas disk being above the threshold for efficient molecular
gas formation, and thus result in higher molecular gas masses and
higher star formation rates, even though the sizes of the gas disks
may not be significantly smaller than at lower redshifts. This may be
relevant given recent observational and theoretical evidence for clumpy
disks at high redshifts (e.g. Elmegreen \etal 2007; 2009; Agertz,
Teyssier, \& Moore 2009; Ceverino, Dekel, \& Bournaud 2009; Dekel,
Sari, \& Ceverino 2009).

\begin{figure*}
\centerline{
\psfig{figure=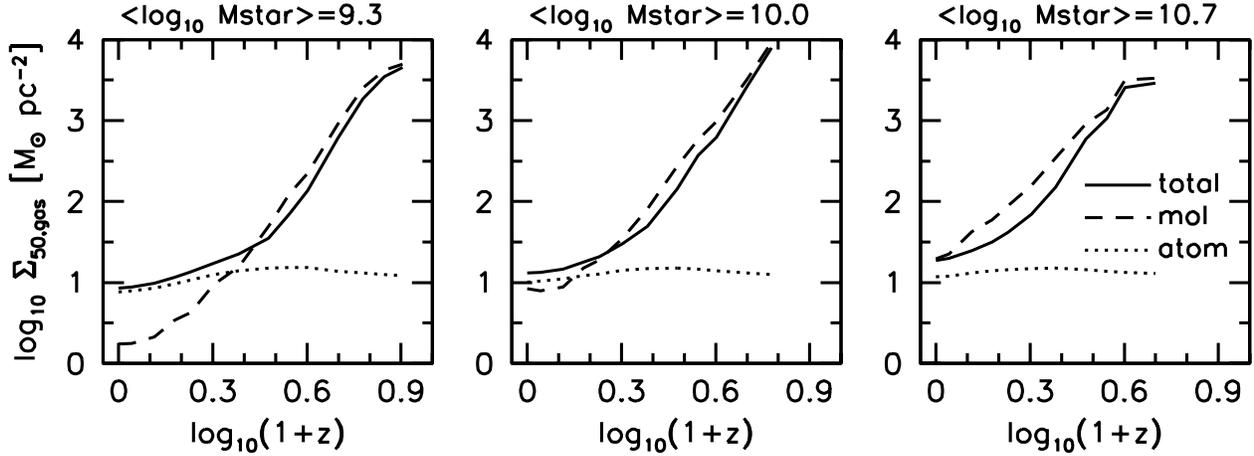,width=0.99\textwidth}}
\caption{ Evolution in the effective gas surface density for the EFB
  model, where the line types correspond to the atomic (dotted),
  molecular (dashed), and total (atomic plus molecular; solid) gas.
  The three panels correspond to three different stellar masses, as
  indicated. Note that the surface densities of atomic gas are
  virtually constant (at $\sim 10 \Msun \pc^{-2}$) while the surface
  densities of molecular gas evolve strongly with redshift. This
  reflects the fact that at surface densities in excess of $\sim 10
  \Msun \pc^{-2}$, the midplane pressure in the disk is sufficiently
  high to assure an efficient conversion of atomic to molecular gas
  (see Appendix~A).}
\label{fig:sgasmz2}
\end{figure*}

\subsubsection{Molecular vs. atomic gas}

We have shown that while the gas fractions show only weak evolution
with redshift, gas densities are much higher at higher
redshift. Hence, one would expect that the {\it molecular} gas
fractions will also be higher at higher redshift.
{Fig.~\ref{fig:gsmxz2}} shows the ratios of total, atomic and
molecular gas to stellar mass for the EFB model as functions of
redshift. As expected, at a fixed stellar mass, the molecular gas
masses increase by an order of magnitude from low to high redshift,
whereas the atomic gas masses decrease by a corresponding
amount. Recent observations of molecular gas masses in star forming
galaxies at redshifts $z\simeq 1.2-2.3$ support this result (Daddi
\etal (2009b); Tacconi \etal 2010). Given that the Schmidt
star formation law for molecular gas has a slope $\simeq 1$
(e.g. Bigiel \etal 2008), this implies that the SFRs are roughly
proportional to molecular gas mass (Eq.4). Thus the increase in
molecular gas fraction provides the fuel for the higher star formation
rates at higher redshifts.

{Fig.~\ref{fig:sgasmz2}} shows the evolution in the effective gas
surface density for the EFB model. The solid line corresponds to the
total gas as in {Fig.~\ref{fig:sgasmz}}. The dashed lines show the
molecular gas density, while the dotted lines show the atomic gas
density. The molecular gas density shows a strong evolution with
redshift, whereas the atomic gas density remains roughly constant at
$\Sigma_{50,\rm gas}\simeq 10 \rm \,\Msun\,pc^2$ all the way from
$z\sim 6$ to $z=0$.

In the near future the Atacama Large Millimeter/submillimeter Array
(ALMA) will have the sensitivity to test our theoretical prediction
that the molecular gas densities and molecular gas fractions of disk
galaxies increase strongly with redshift. A test of our predictions
for the evolution of the mass fractions and densities of atomic gas,
however, most likely will have to wait for the Square Kilometer Array
(SKA), which is currently scheduled to start operation no earlier than
2018.  In the meantime, measurements of \hi column densities from
damped Ly$\alpha$ systems (DLAs) provide a means of measuring the
atomic gas content of high redshift galaxies. The analysis of 738 DLAs
by Prochaska \& Wolfe (2009) suggests that \hi disks have not evolved
significantly since $z\sim 2$. Although in agreement with our model
predictions, a more detailed comparison between the observed DLA
column densities and our model predictions is needed to assess whether
they are indeed consistent with each other.

In summary, the increase in {\it net} gas accretion rate from low to
high redshift provides the means for the SFRs to increase at a fixed
stellar mass. The mechanism for converting this elevated gas accretion
rate into an elevated star formation rate is an increase in gas
density, not an increase in gas mass. Increased gas density, mainly
due to the fact that haloes are denser at higher redshifts, results in
increased molecular gas fractions, which provides the increased fuel
supply for the higher SFRs at high redshifts.

\begin{figure*}
\centerline{
  \psfig{figure=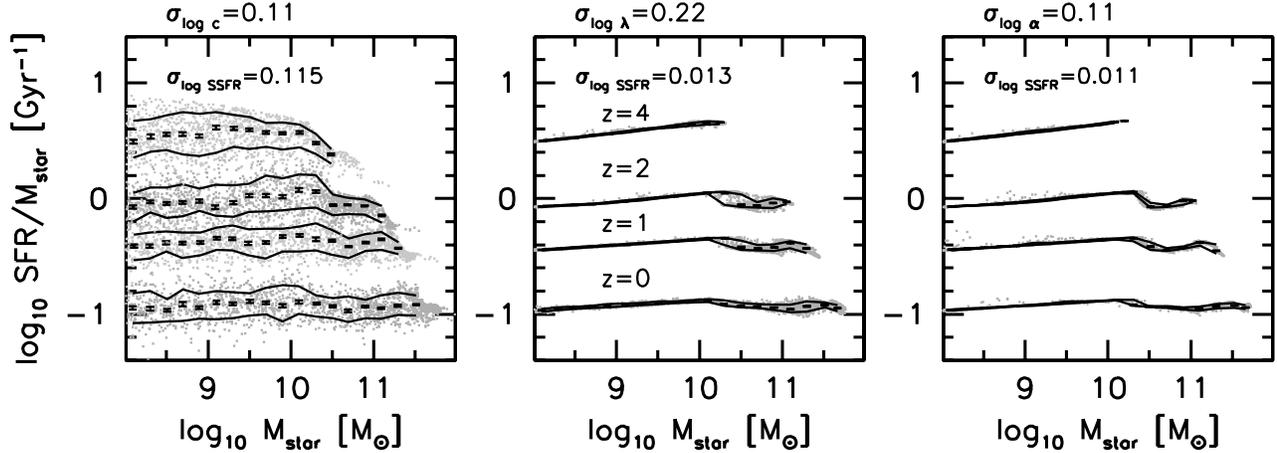,width=0.99\textwidth}}
\caption{The impact of different sources of scatter on the scatter in
  the SFR sequence for the EFB model: halo concentration (left panel);
  halo spin parameter (middle panel); halo AMD parameter (right
  panel).  The model galaxies at redshifts $z=0,1,2,$ \& $4$ are
  indicated by grey points, the black error bars show the error on the
  median SSFR in $\Mstar$ bins of 0.2 dex width, and the black lines
  enclose 68.3\% of the model galaxies.  The average scatter in SSFR
  over the four redshifts is given in each panel. The scatter in SSFRs
  is clearly dominated by scatter in the halo concentrations, while
  scatter in the spin parameter or AMD parameter result in negligible
  scatter in SSFRs.}
\label{fig:sfrm-efb-cla}
\end{figure*}

\begin{figure*}
\centerline{
\psfig{figure=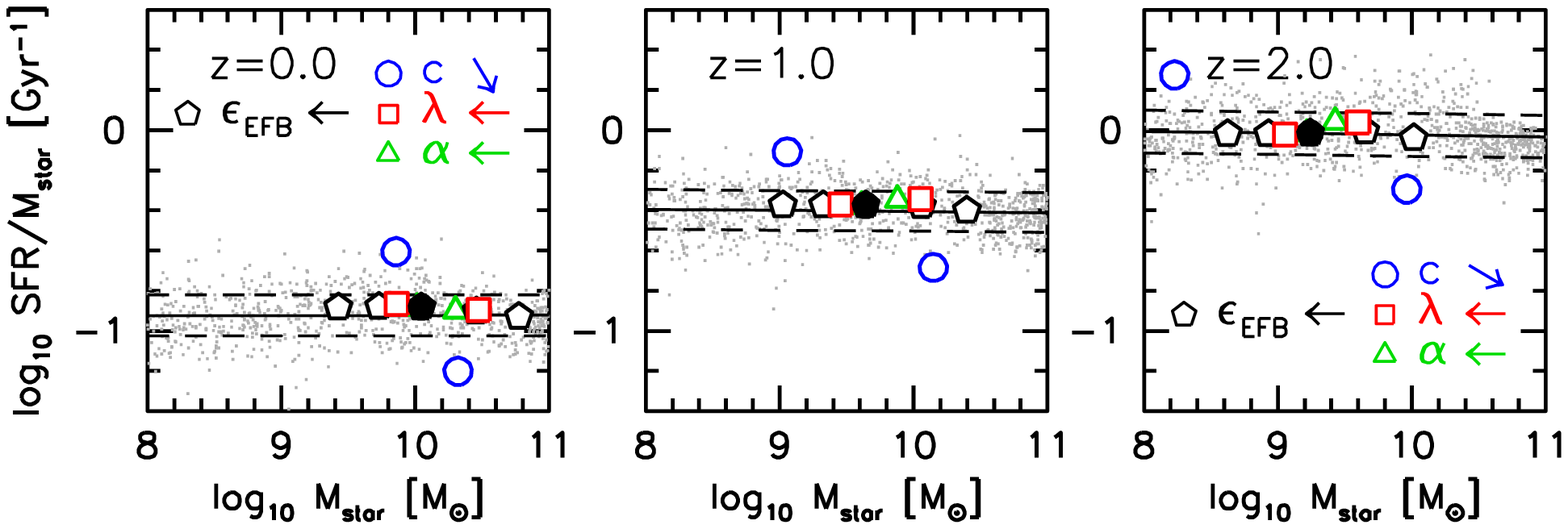,width=0.99\textwidth}}
\centerline{
\psfig{figure=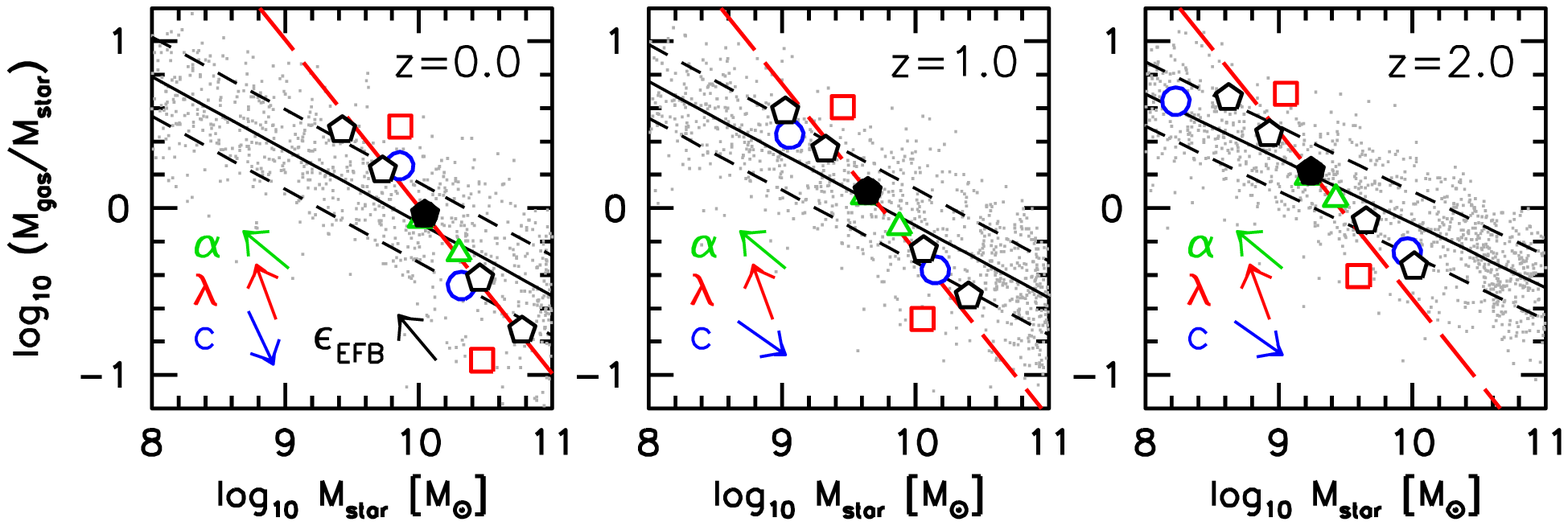,width=0.99\textwidth}}
\centerline{
\psfig{figure=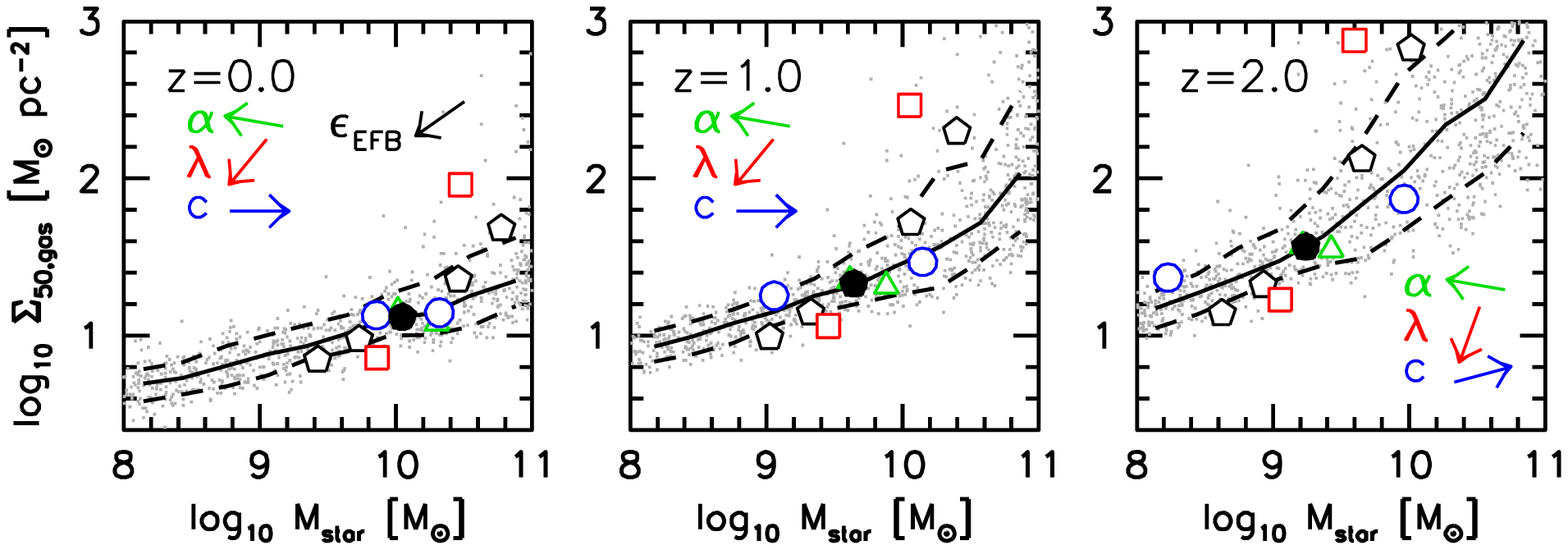,width=0.99\textwidth}}
\caption{ Contribution of variation in feedback efficiency (black
  pentagons) and scatter in halo concentration, $c$ (blue circles),
  halo spin, $\lambda$ (red squares), and halo AMD, $\alpha$ (green
  triangles), to the SFR sequence (SSFR vs. $\Mstar$, upper panels),
  the gas-to-stellar mass ratio - stellar mass relation (middle
  panels), and the gas effective surface density - stellar mass
  relation (lower panels). The red long-dashed line in the middle
  panels corresponds to constant gas mass. The grey points show the
  EFB model with scatter in all three parameters at redshifts $z=0$
  (left column), $z=1$ (middle column), and $z=2$ (right column).  The
  big symbols show models with $M_{\rm vir,0}=4.5 \times 10^{11}
  \Msun$ and median halo parameters (solid black), and $\pm 2 \sigma$
  scatter (open symbols). The arrows indicates the direction the
  galaxy moves when the parameter in question increases. SFR,
  gas-to-stellar mass ratio, gas surface density and $\Mstar$ depend
  mostly on $c$ and $\lambda$. The scatter in the gas-to-stellar mass
  ratio and the gas surface density are dominated by $\lambda$, but
  $\lambda$ moves galaxies along the SFR sequence, and thus does not
  contribute to its scatter.}
\label{fig:sfrm-efb-cla2}
\end{figure*}

\section{Origin of scatter in the SFR sequence}

In this section we address the question of why the scatter in the SFR
sequence is so small. In our model there are 3 sources of scatter:
halo concentration, $c$, halo spin, $\lambda$, and halo AMD,
$\alpha$. We assume the scatter in these three parameters are
independent of each other, and independent of redshift.

{Fig.~\ref{fig:sfrm-efb-cla}} shows the SFR sequence (expressed in
terms of SSFR vs. $\Mstar$) at redshifts $z=0,1,2$, and $4$ for the
EFB model. Each model has only one source of scatter: halo
concentration (0.11 dex) halo spin (0.22 dex), halo AMD (0.11 dex).
This shows that the scatter in the SFR sequence, at all redshifts is
dominated by the scatter in halo concentration, and rather remarkably,
shows only a very weak dependence on halo spin or AMD.

\subsection{Why does the scatter in the SFR sequence depend on halo
  concentration?}
\label{ssfr:c}
In our model the concentration of the dark halo is directly coupled to
the mass accretion history (MAH) of the dark matter halo (see DB09 for
details).  In order to assess the relative importance of scatter in
MAH vs. scatter in halo concentration, we have also run models in
which the concentration is decoupled from the MAH. These show that
scatter in the halo concentration parameter contributes only a small
amount to the scatter in the SFR sequence. Thus it is mainly the
variations in the MAHs that dominates the scatter in the SFR sequence,
at least in our models.  Higher concentration halos have MAHs shifted
to earlier times (i.e., these haloes assemble their mass earlier), and
vice versa for lower concentration halos.  An earlier MAH results in
higher baryon accretion rates at high redshifts, which in turn implies
higher SFRs, but also a larger stellar mass.

In our model the scatter in MAH is determined by scatter in halo
concentration, which is 0.14 dex of all haloes (Bullock \etal 2001a)
and just 0.11 dex for relaxed haloes (Wechsler \etal 2002; Macci{\`o}
\etal 2007). Since we adopt a smooth MAH, and we assume a one-to-one
correspondence between halo concentration and formation redshift, it
is likely that our model underestimates the true scatter in baryon
accretion rates onto central galaxies. Determining the true scatter
requires cosmological hydrodynamical simulations. The simulations used
in Dekel \etal (2009) find a scatter of about 0.3 dex in the gas
accretion rates onto galaxies in haloes of mass $10^{12}\Msun$ at
$z=2.5$. Assuming this scatter translates linearly into scatter in
SSFR (as in our models) this would reproduce the full observed scatter
in the SFR sequence. Further studies are needed to quantify the mass
and redshift dependence of this scatter.

The upper panels of {Fig.~\ref{fig:sfrm-efb-cla2}} show the effect of
varying halo concentration $c$ (and thus also the MAH), halo spin
parameter, $\lambda$, and AMD, $\alpha$, on the SFR sequence. The
solid black circle shows a model with median $c$, $\lambda$ and
$\alpha$, and with energy driven feedback with $\epsilon_{\rm
  FB}=0.25$. The open blue circles show models with $\pm 2\sigma$
variation in $c$. The direction the model moves for increasing $c$ is
given by the blue arrows. This shows that galaxies that form in haloes
with higher concentrations (i.e. earlier MAHs) lie below the SFR
sequence (i.e. lower SFR at fixed $\Mstar$), while galaxies that form
in haloes with lower concentrations lie above the SFR sequence
(i.e. higher SFR at fixed $\Mstar$).  This implies that the
normalization of the SFR sequence depends on the cosmological
parameters, in particular $\sigma_8$, with higher $\sigma_8$ resulting
in lower SFR at a given $\Mstar$ (see \S~\ref{sec:cos})

\begin{figure*}
\centerline{
  \psfig{figure=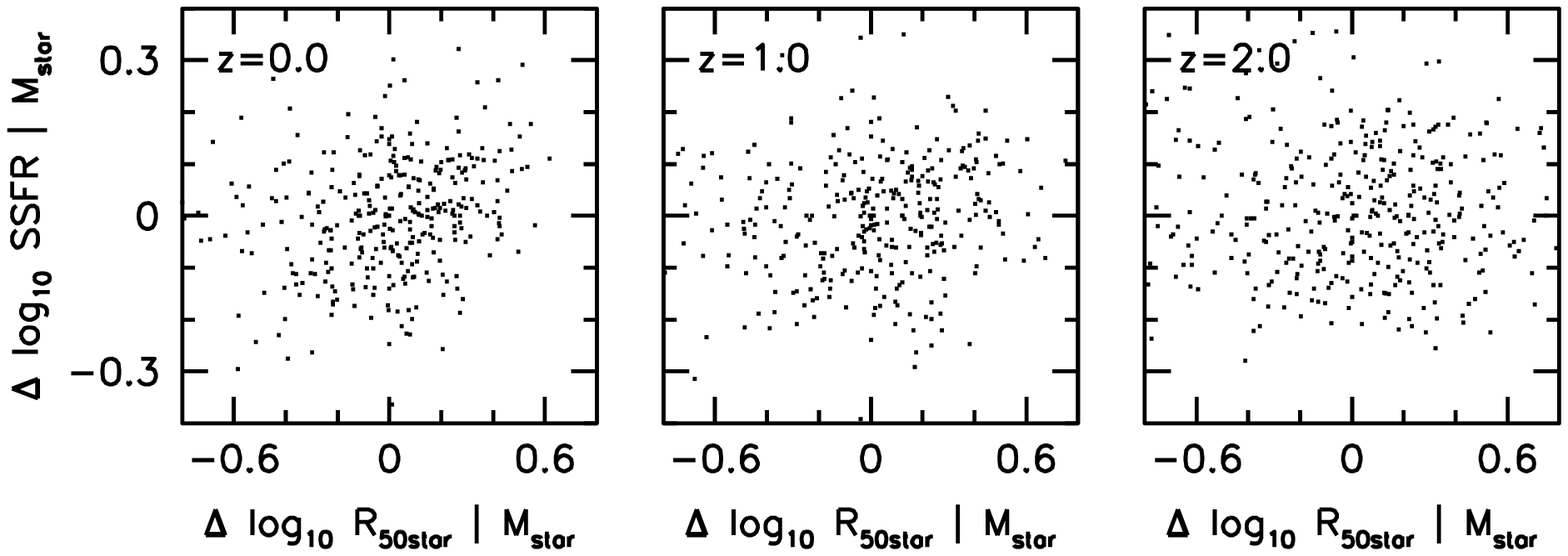,width=0.99\textwidth}}
\centerline{
  \psfig{figure=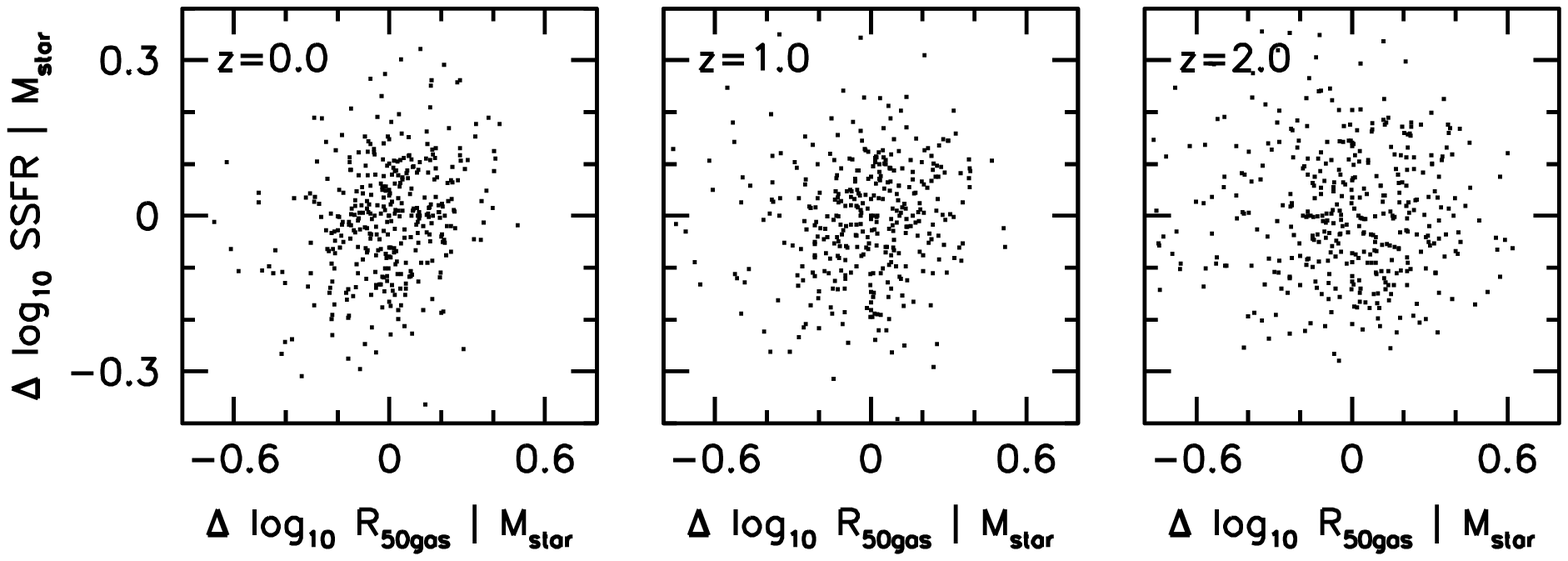,width=0.99\textwidth}}
\caption{ Correlation between the scatter in specific star formation
  rate (SSFR) at fixed stellar mass to scatter in disk size at fixed
  stellar mass in the EFB model at redshifts $z=0,1,$ \& $2$. The disk
  sizes are stellar half mass radii (upper panels) and cold gas half
  mass radii (lower panels). This shows that the scatter in the SFR
  sequence is independent of disk size (both stellar and gaseous), as
  expected from the independence of the SFR sequence to spin
  parameter.}
\label{fig:dssfrdr}
\end{figure*}

\subsection{Why is the scatter in the SFR sequence independent of halo
  spin?}

The red open squares in {Fig.~\ref{fig:sfrm-efb-cla2}} show models
with $\pm 2\sigma$ variation in $\lambda$, while the green open
triangles show models with $\pm 2\sigma$ variation in $\alpha$.  As
for the concentrations, the direction the model moves for increasing
$\lambda$ and $\alpha$ is indicated by the red and green arrows,
respectively.  This shows that both SFR and $\Mstar$ depend on the
halo angular momentum parameters, and spin in particular, but that the
changes move galaxies {\it along} the SFR sequence: lower $\lambda$
and $\alpha$ both result in higher SFR and higher $\Mstar$.

The middle and lower panels of {Fig.~\ref{fig:sfrm-efb-cla2}} show the
effect of variation in $c$, $\lambda$, and $\alpha$, on the relation
between stellar mass and the gas-to-stellar mass ratio (middle panels)
and between stellar mass and gas density (lower panels).  The scatter
in these two relations is dominated by $\lambda$, with $c$ a
sub-dominate component, and $\alpha$ a marginal component.  Changes in
$\alpha$ (AMD) only have a very small effect, which explains why in
many cases you can only see one green triangle (that for low values of
$\alpha$).  Smaller values for $\lambda$ result in higher gas
densities, which result in more efficient star formation, and hence
higher SFRs, higher $\Mstar$, and lower $\Mgas$. This explains why the
SFR sequence is invariant under changes in $\lambda$.

\subsection{Independence of the SFR sequence to disk size} In our
model the scatter in disk sizes, at fixed stellar mass, is dominated
by scatter in the spin parameter. Scatter in halo concentration (which
is coupled to scatter in MAH) also contributes, but to a much lesser
extent (see for example Dutton \etal 2007, Fig.7).  Thus we would
expect that the SFR sequence is also independent of disk size. In
Fig.~\ref{fig:dssfrdr} we show that this is indeed the case.  This
figure shows the scatter in SSFR at fixed stellar mass vs. the scatter
in disk size at fixed stellar mass, for galaxies in the EFB model with
stellar masses between $9.0 < \log_{10} \Mstar < 10.5$. We get similar
results, i.e. no correlation, for our other feedback models.  In the
upper panels the disk size is the stellar half mass radius, while in
the lower panels the disk size is the cold gas half mass radius.  The
left panels show results at redshift $z=0$, the middle panels $z=1$
and the right panels $z=2$.

We now return to the line of reasoning from the introduction, that
suggested that scatter in disk size would be expected to result in
scatter in the SFR sequence, and explain why it is false (at least in
our model).  Recall that for a SK star formation law with $n=1.5$ the
SFR depends linearly on gas mass and gas disk size. Thus for fixed gas
mass, one expects that galaxies with smaller gas disks should have
higher SFRs. If stellar mass is proportional to gas mass, and stellar
disk sizes are proportional to gas disk sizes, then one would expect
that at fixed stellar mass, scatter in stellar disk sizes to result in
scatter in SFRs. The reason that this is not the case in our models is
that the scatter in the relation between gas mass and stellar mass
depends on disk size. At a fixed stellar mass, galaxies with smaller
disks have lower gas fractions, and galaxies with larger disks have
higher gas fractions. These two effects cancel out, leaving the
SFR unchanged. 

Another way to think about this is that at fixed halo mass there is a
spread in disk sizes. Galaxies with smaller disks do have higher SFR,
but at the same time they also have higher stellar masses (and lower
gas fractions). Thus scatter in disk sizes creates a SFR sequence at
fixed halo mass. 

\begin{figure*}
\centerline{
  \psfig{figure=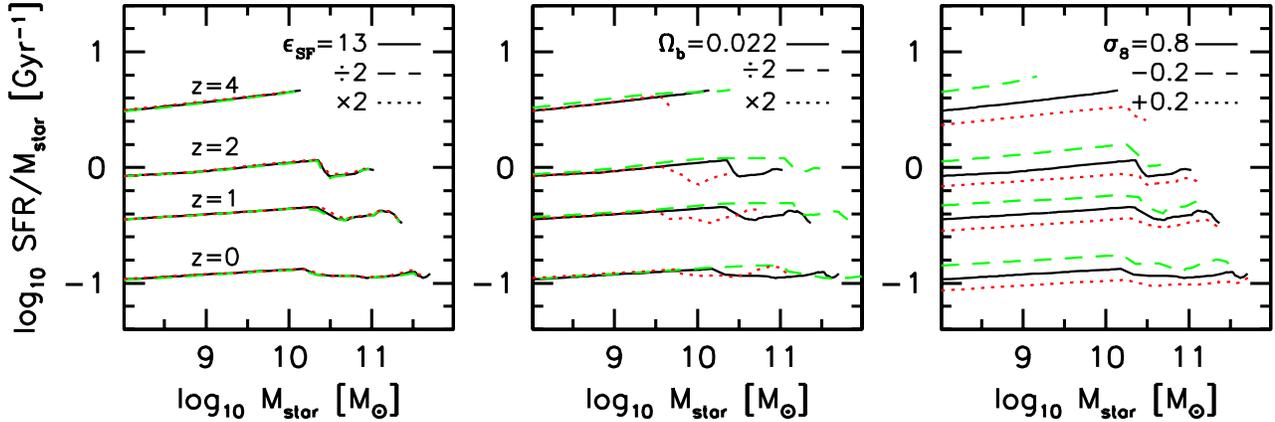,width=0.99\textwidth}}
\caption{Dependence of the specific star formation rate (SSFR) -
  stellar mass relation on the star formation efficiency parameter,
  $\tilde{\epsilon}_{\rm SF}$ (left panel); cosmological baryon
  density, $\Omega_{\rm b}$ (middle panel); and the cosmological power
  spectrum amplitude, $\sigma_8$ (right panel). The relation due to
  our adopted EFB model is given by the solid black lines. Relations due
  to lower/higher values of these parameters are given by dashed green
  and dotted red lines, respectively. The SSFR - \Mstar relation is
  independent to changes in the star formation efficiency parameter
  (at the factor of 2 level), and also largely independent of the
  cosmological baryon density. The zero point of the SSFR-\Mstar
  relation is dependent on $\sigma_8$, as this parameter controls the
  timescale of the halo mass accretion history.  }
\label{fig:ssfrm-misc}
\end{figure*}

The lack of a correlation we find between scatter in SSFR and scatter
in disk size (both stellar and gas) at fixed stellar mass should be
testable using existing and future data. Indeed, correlations between
SSFR and stellar surface density have been studied by Franx \etal
(2008) and Williams \etal (2009). These authors have shown that there
is a strong correlation between SSFR and stellar surface density, such
that galaxies with higher stellar densities have lower SSFR.  Williams
\etal (2009) also studied the galaxy half light radius - stellar mass
relation as a function of SSFR from redshifts $z=0$ to $z=2$. At a
fixed stellar mass, there is a clear trend for smaller galaxies to
have lower SSFR. These observations may seem at odds with the
predictions from our models. However, our predictions are for galaxies
on the SFR sequence, and not for galaxies in general.  Thus it is
plausible that the stellar surface density is important for
determining whether a galaxy is on or off the SFR sequence. But that
for galaxies on the SFR sequence, the stellar density does not
correlate with the SSFR, as in our models.  Another caveat is that the
sizes measured by Franx \etal (2008) and Williams \etal (2009) are
global half light sizes, whereas our model sizes are disk
sizes. Global sizes do not distinguish between bulges and disks. Since
disks tend to be larger than bulges as well as having higher SSFRs, a
variation of bulge fraction at fixed stellar mass will naturally
result in a correlation between galaxy half light radius and SSFR,
even if no such correlation exists between disk size and SSFR.  Thus
averaging bulges and disks together may be hiding some important clues
to the strength and nature of the correlation between SSFR and stellar
surface density. Future observational studies will be able to address
these issues.

\section{Robustness of the SFR sequence}
\label{sec:fb}

\subsection{Dependence on feedback }
The black open pentagons in the upper panels of
{Fig.~\ref{fig:sfrm-efb-cla2}} show the effect of varying the energy
feedback efficiency, $\epsilon_{\rm EFB}$, from 0 to 1 on the SFR
sequence ($\epsilon_{\rm EFB}$ increases from right to left as
indicated by the black arrow).  Varying $\epsilon_{\rm EFB}$ moves
galaxies along the SFR sequence, in the sense that models with
stronger feedback have both lower stellar masses and lower SFRs, but
the same SSFR. Increased feedback efficiency results in lower gas
densities (lower panels), and hence less efficient star formation
(upper panels), and higher gas-to-stellar mass ratios (middle panels).

The fact that feedback results in higher cold gas fractions might seem
counter-intuitive, as feedback blows cold gas out of the system which
lowers the cold gas to halo mass fraction.  However, the outflows in
our model occur over an extended period of time, and thus the galaxy
has time to adjust to them. Outflows reduce the baryonic masses, while
at the same time increasing the specific angular momentum of the
remaining baryons (Maller \& Dekel 2002; Dutton \& van den Bosch
2009). These two effects cause the density of the gas disk to
decrease, resulting in less efficient star formation and hence higher
cold gas fractions for the remaining baryons. In fact, it is
interesting to note that varying the feedback efficiency moves
galaxies roughly along a line of constant gas mass (red long-dashed
line in {Fig.~\ref{fig:sfrm-efb-cla2}}).

\begin{figure*}
\centerline{
\psfig{figure=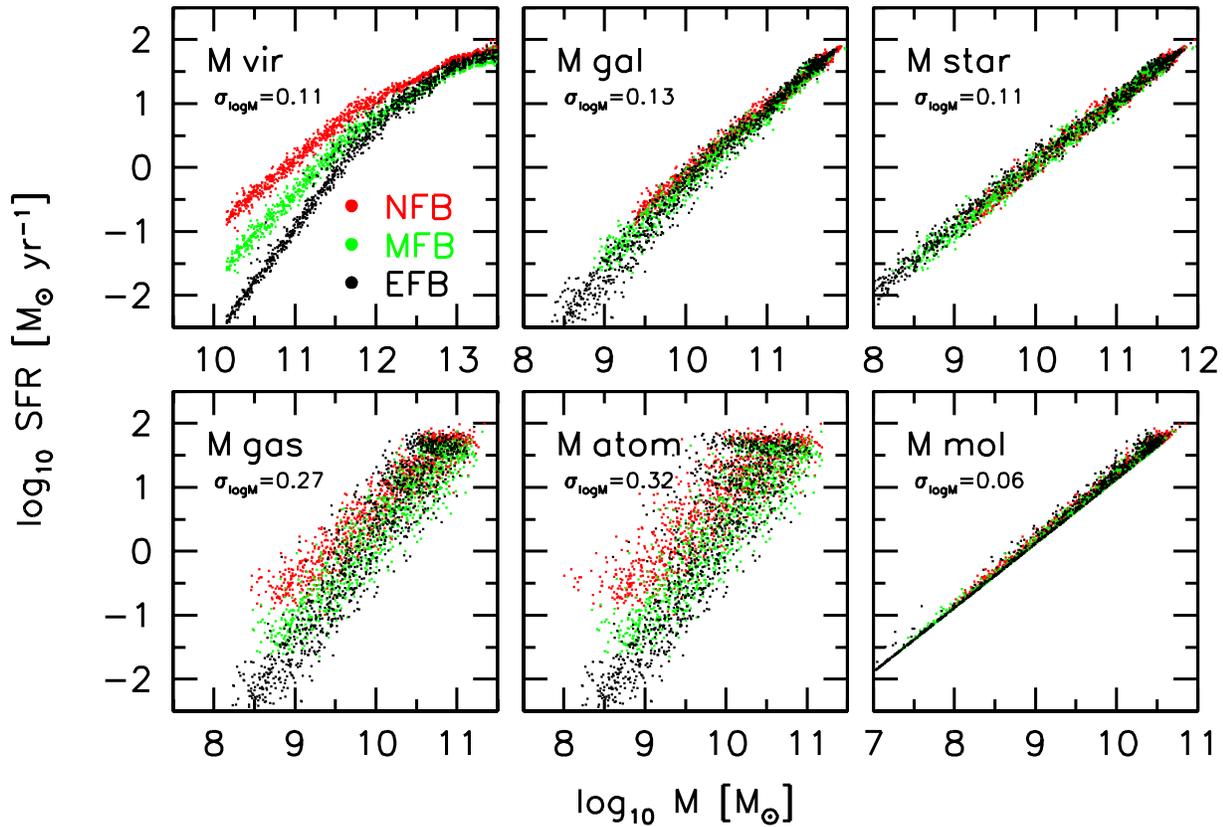,width=0.95\textwidth}}
\caption{Various relations between SFR and mass at $z=0$ for three
  feedback models: no feedback (NFB, red dots), momentum driven
  feedback (MFB, green dots); and energy driven feedback (EFB, black
  dots). Results are shown for six different `types' of mass: total
  virial mass ($\Mvir$, upper left); galaxy mass ($\Mgal$, upper
  middle); stellar mass ($\Mstar$, upper right); cold gas mass
  ($\Mgas$, lower left); cold atomic gas ($\Matom$, lower middle); and
  cold molecular gas ($\Mmole$, lower right). The tightest relation is
  that between SFR and $\Mmole$, and the weakest relation is that
  between SFR and $\Matom$. The relations between SFR and $\Mvir$,
  $\Mgas$, and $\Matom$ are all sensitive to the feedback model,
  whereas those between SFR and $\Mgal$, $\Mstar$, $\Mmole$, are
  insensitive to the feedback model.}
\label{fig:sfrmall}
\end{figure*}

\subsection{Dependence on star formation efficiency}
One of the uncertainties in our star formation law is the
normalization, which depends on amongst other things, the stellar IMF.
The right panel of Fig.~\ref{fig:ssfrm-misc} shows that the SFR
sequence is independent to changes in the star formation efficiency
parameter ($\tilde{\epsilon}_{\rm SF}$) by a factor of 2.
Furthermore, for individual galaxies, SFR and \Mstar are insensitive
to $\tilde{\epsilon}_{\rm SF}$. This is because changes (e.g. an
increase) in $\tilde{\epsilon}_{\rm SF}$ are offset by changes (e.g. a
decrease) in gas masses and gas densities, i.e. star formation is
self-regulating.  Thus differences in IMF, such as between Salpeter
and Chabrier (2002), do not impact on the SFR sequence in our models.

\subsection{Dependence on cosmological parameters}
\label{sec:cos}
As discussed in \S \ref{ssfr:c} the SFR sequence depends on the MAH of
the baryons, which is cosmology dependent. The MAH can be described to
lowest order by a normalization, and a formation redshift.

We simulate variation in the normalization by changing the
cosmological baryon fraction by plus or minus a factor of 2. As shown
in the middle panel of Fig.~\ref{fig:ssfrm-misc} this has no effect on
the normalization of the SSFRs. The only effect is to change the mass
scale at which cooling becomes inefficient, which occurs because
changing the baryon fraction changes the density of the hot halo gas,
and hence the cooling efficiency.

We simulate variation in the formation redshift of the haloes by
varying the normalization of the power spectrum, $\sigma_8$. Higher
$\sigma_8$ results in earlier forming haloes, and hence an earlier
MAH. The right panel of Fig.~\ref{fig:ssfrm-misc} shows the impact on
the SFR sequence of varying $\sigma_8$ from 0.6 to 1.0. The SSFRs are
higher for lower $\sigma_8$, and lower for higher $\sigma_8$.

Thus the zero point and evolution of the SFR sequence appears to
depend only on the time dependence of the gas mass accretion
history. This is a result of galaxies being in a quasi steady state
between gas accretion rate and star formation rate, which in turn is
the result of the self-regulating nature of star formation in galaxy
disks.
This suggests that in order to change the slope or time evolution of
the SFR sequence relation, it is necessary to change the gas accretion
history, rather than the details of how gas is converted into stars.

\subsection{Is the SFR sequence special?}

We have shown that the SFR sequence in our models is independent of
the feedback efficiency, and has small scatter.  We now consider
relations between the SFR and other masses, to see if the SFR sequence
is special.

{Fig.~\ref{fig:sfrmall}} shows the SFR - $M$ relations at redshift
$z=0$, where $M$ is the virial mass ($\Mvir$), galaxy mass
($\Mgal=\Mstar+\Mgas$), stellar mass ($\Mstar$), cold gas mass
($\Mgas=\Matom+\Mmole$), cold atomic gas ($\Matom$), and cold
molecular gas ($\Mmole$). In each panel results are shown for all
three models; NFB, EFB and MFB (as indicated). As we have already
seen, the SFR sequence (upper right panel) has small scatter (0.11 dex
in $\Mstar$ at SFR $= 0.1$ to $10 \,\rm\Msun\,yr^{-1}$) and a slope
that is independent of mass and feedback. The SFR - $\Mgal$ relation
(upper middle panel) has a slightly larger scatter (0.12 dex), which
increases at the low mass end, and has a slope that is weakly
dependent on mass and feedback. The SFR - $\Mvir$ relation (upper left
panel) has a slightly smaller scatter (0.10 dex) than the SFR
sequence, but the slope is strongly dependent on mass and the feedback
model. The SFR - $\Mgas$ relation (lower left panel) has a large
scatter (0.26 dex), and a slope that depends on the feedback
model. The SFR - $\Matom$ relation (lower middle panel) has the
largest scatter (0.32 dex) and also has a slope that depends on
feedback. Finally, the SFR - $\Mmole$ relation (lower right panel) has
the smallest scatter (0.06 dex) and is independent of feedback.

The origin of the tight relation between SFR and $\Mmole$ is due to
our effective star formation law, $\Sigma_{\rm SFR} \propto
\Sigma_{\rm mol}^n$, with a Schmidt-law index of $n\simeq 1$. Thus from
Eq.(4), the integrated SFR of a galaxy (which reflects its stellar
mass) is proportional to the total molecular gas mass.


\section{Summary}
\label{sec:conclusion}

We have used a disk galaxy evolution model to investigate the origin
of the zero point evolution and small scatter of the star formation
rate sequence --- the tight correlation between galaxy star formation
rate (SFR) and stellar mass ($\Mstar$). In our model the mass
accretion histories of the dark matter haloes are smooth (i.e. there
are no mergers), gas accretes onto the central galaxy via ``hot mode''
accretion (i.e. gas that enters the halo shock heats to the virial
temperature, cools radiatively, and then accretes onto the central
galaxy in a free fall time), the cold gas density profile is
determined by detailed conservation of specific angular momentum, the
radially dependent molecular gas fraction is taken to depend on the
mid-plane pressure in the disk, and star formation occurs at a rate
that is governed by the local surface density of the molecular gas.
We summarize our results as follows:

\begin{itemize}

\item The slope, zero point and scatter of the SFR sequence are
  independent of our feedback model.  This is because feedback effects
  the SFRs and $\Mstar$ such that galaxies are shifted along the
  relation, rather than perpendicular to it.

\item The zero point of the SFR sequence in our models evolves
  strongly, and in good agreement with observations between $z\simeq
  4$ to $z=0$. However, the models under predict the SFRs at $z \sim
  2$, in qualitative agreement with the hydrodynamical simulations of
  Dav{\'e} (2008), who argued for evolution in the IMF in order to
  reconcile his simulations with the data. However, our model fits the
  data at $z \simeq 4$, disfavoring such an evolving IMF model.

\item In our model the evolution in the zero point of the SFR sequence
  closely follows the evolution in the gas accretion rate.  However,
  at a fixed stellar mass the absolute value of specific gas accretion
  rate depends strongly on the feedback model.  There is a better
  correlation between the star formation rate and the {\it net} gas
  inflow rate (inflow - outflow).  The high (central) densities of
  \LCDM haloes and the relatively low values of the spin parameter
  aspire to assure that the majority of (centrifugally supported)
  disks always attain sufficiently high gas densities for star formation
  to be efficient, and thus for a steady state to exist in which the
  star formation rate follows the {\it net} inflow rate of cold gas.

\item Our models predict that the SSFR should decrease monotonically
  from redshifts $z\simeq 7$ to $z\simeq 2$, whereas the current
  observations favor no evolution over this redshift range. At $z\simeq
  7$ the discrepancy between models and observation is a factor of
  $\simeq 4$. Thus, if the observations are correct, this indicates
  that the star formation rate no longer follows the gas accretion
  rate in high redshift ($z\gta 2$) galaxies, as we infer that it does
  in low redshift ($z\lta 2$) galaxies.

\item At a fixed stellar mass, gas masses evolve only very weakly in
  our model, whereas gas densities evolve strongly. Thus the higher
  SSFRs in higher redshift galaxies is due to higher gas densities,
  and perhaps surprisingly, {\it not} due to higher total cold gas
  masses (as is often claimed e.g. Daddi \etal 2009b). The higher gas
  densities are driven by the higher densities of \LCDM haloes at
  higher redshifts coupled to a decrease in disk specific angular
  momentum at higher redshifts.

\item The increased gas densities at higher redshifts result in higher
  molecular gas masses. At a fixed stellar mass, the molecular gas
  masses are a factor of $\simeq 10$ higher at $z\simeq 3$ compared to
  $z=0$. This thus provides the increased fuel for the higher SSFRs in
  high redshift galaxies.

\item At fixed stellar mass, the molecular gas densities in our model
  increase strongly with increasing redshift, whereas the atomic gas
  densities are roughly independent of redshift, with an effective \hi
  surface density of $\Sigma_{50,{\rm HI}} \simeq 10
  \,\rm\Msun\,pc^{-2}$. This is basically a reflection of the fact
  that at \hi surface densities higher than this, the atomic gas is
  efficiently converted into molecular gas (see also Martin \&
  Kennicutt 2001; Wong \& Blitz 2002; Bigiel \etal 2008).

\item The scatter in SFR at fixed $\Mstar$ in our models is $\simeq
  0.12\pm0.01$ dex.  The source of this scatter is the variance in
  mass accretion histories for haloes of a given virial mass, which in
  our model is coupled to the halo concentration. Scatter in halo spin
  contributes negligibly, because spin scatters galaxies along the SFR
  sequence. Note that the $0.12 \pm 0.01$ dex scatter in our models is
  less than the observed scatter ($\sim 0.3$ dex), as it should be
  given that the observed scatter is likely to be dominated by
  observational uncertainties. However, it is likely that our model
  underestimates the scatter due to our simplified treatment of the
  halo mass accretion history. It remains to be determined whether the
  true intrinsic scatter is indeed as small as predicted by our
  models, or whether additional sources of intrinsic scatter are
  required.

\item Relations between SFR and other masses (halo virial mass, galaxy
  mass, gas mass, atomic gas mass) exist, but have either larger
  scatter, or a slope that is dependent on mass and/or the feedback
  model. The tightest correlation is between SFR and molecular mass,
  which is a manifestation of the approximately linear relation
  between SFR surface density and molecular gas surface density which
  is built into our models.

\end{itemize}

While our model is conceptually very simple, it reproduces the main
features of the observed SFR sequence: strong zero point evolution,
small scatter, and a slope that is close to unity. In detail, though,
there seem to be three potential failures: (i) the model slope appears
somewhat {\it too steep} at $z=0$, (ii) the model {\it underpredicts}
the SSFRs at $z\sim 2$, and (iii) the model {\it overpredicts} the
SSFRs at $z\gta4$.  Better data is required in order to asses the
severity of these potential shortcomings.

A key prediction of our model that will soon be testable
observationally (with ALMA) is that, at fixed stellar mass, the
molecular gas densities and molecular gas masses should be
substantially (by at least an order of magnitude) higher at $z \gta 2$
than they are for present-day disk galaxies. Results for small numbers
of disk galaxies at $z\simeq 1-2$ lend support to our predictions
(Daddi \etal 2009b; Tacconi \etal 2010), though these
results need to be verified with larger samples and more robust
measurements.

\section*{Acknowledgements} 
A.A.D thanks Sandra Faber for asking the question that started this
paper.  A.A.D.  acknowledges financial support from the National
Science Foundation grant AST-0808133, from Hubble Space Telescope
grant AR-10965.02-A, and from a CITA National Fellowship.
A.D. acknowledge support from an ISF grant, from GIF I-895-207.7/2005,
from German-Israeli Project Cooperation grant STE1869/1-1.GE625/I5-1,
from France-Israel Teamwork in Sciences, from the Einstein Center at
HU, and from NASA ATP NAG5-8218 at UCSC.


\appendix

\section{Overview of our Star Formation Law} 
\begin{figure*}
\centerline{
\psfig{figure=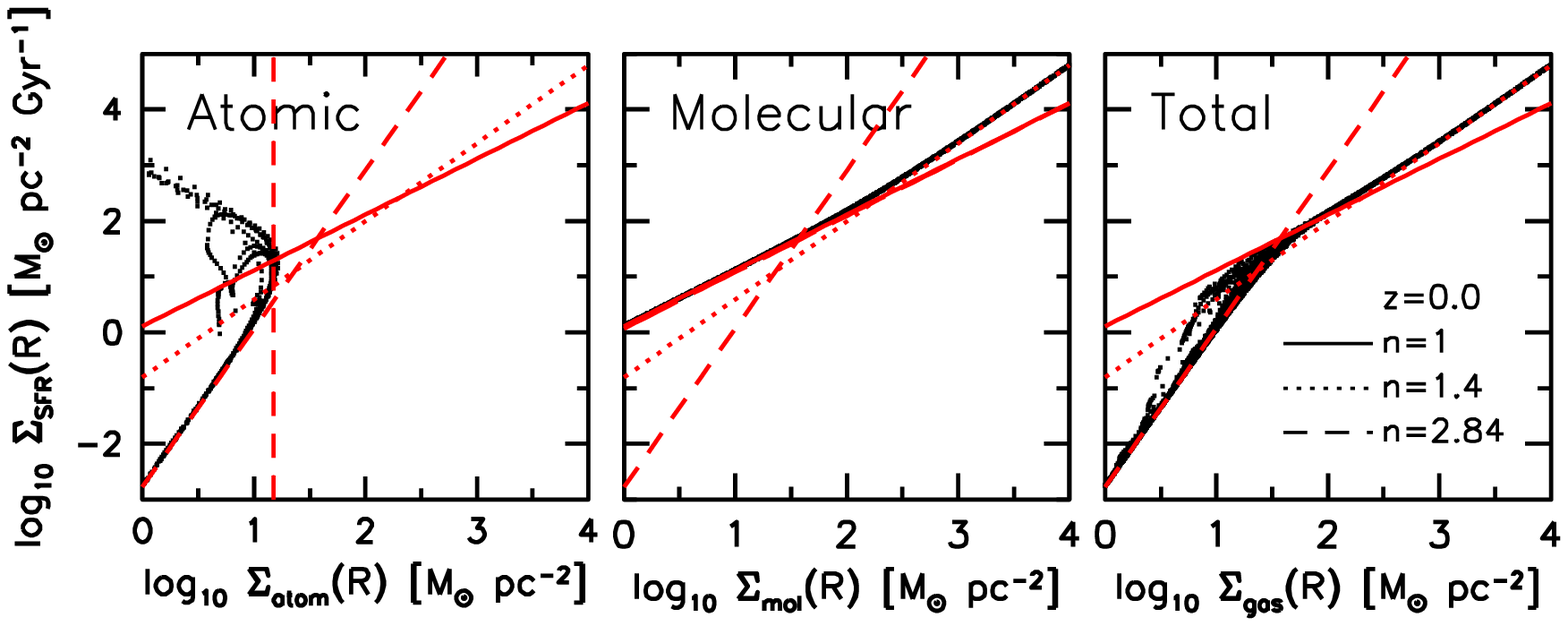,width=0.95\textwidth}}
\centerline{
\psfig{figure=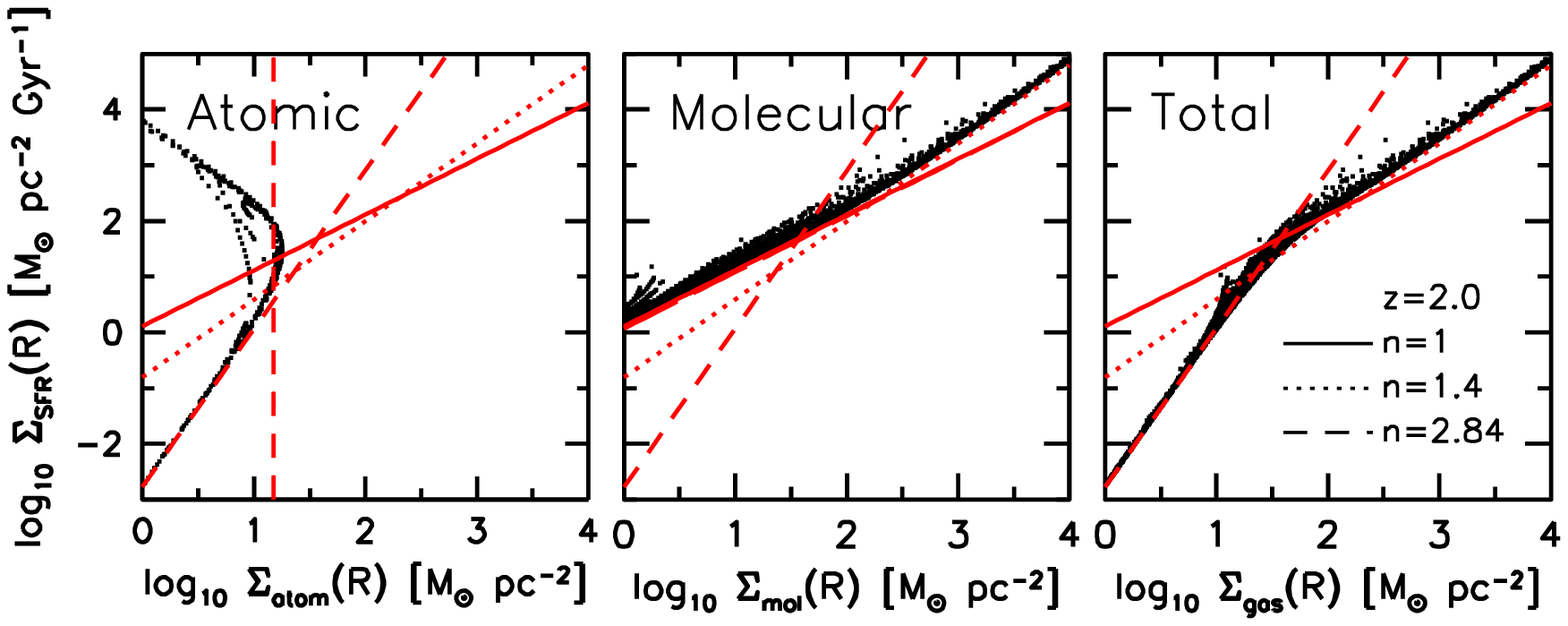,width=0.95\textwidth}}
\centerline{
\psfig{figure=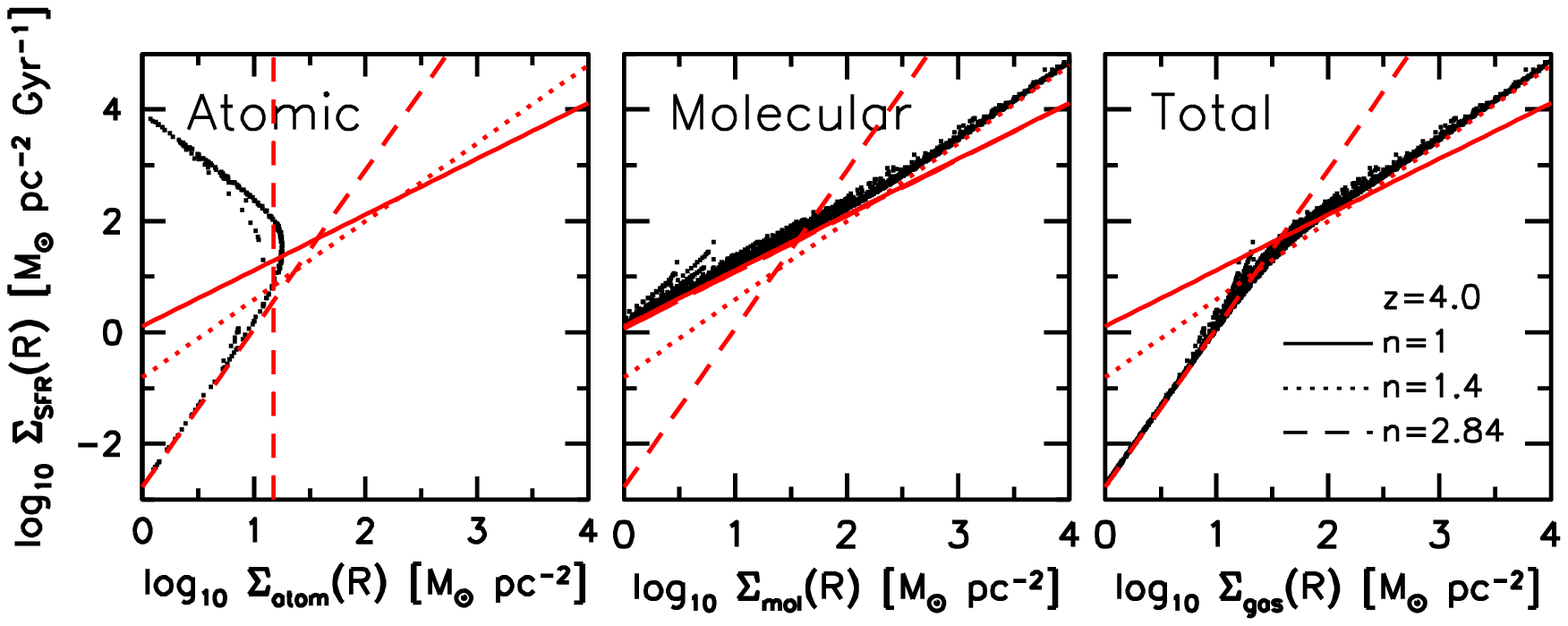,width=0.95\textwidth}}
\caption{ Radial star formation rate surface density ($\Sigma_{\rm
    SFR}(R)$) - gas surface density relations for 100 galaxies from
  our EFB model at redshifts $z=0$ (upper panels); $z=2$ (middle
  panels) and $z=4$ (lower panels). Each dot shows the star formation
  rate surface density and gas surface density in a single radial
  bin. Each galaxy has 10's of such radial bins. The three panels at
  each redshift correspond to the radial: atomic gas surface density
  ($\Sigma_{\rm atom}(R)$, left panels); molecular gas surface density
  ($\Sigma_{\rm mol}(R)$, middle panels); and total gas surface
  density ($\Sigma_{\rm gas}(R)$, left panels). The red lines show the
  theoretical asymptotic relations expected in the high pressure
  ($\Sigma_{\rm gas} \simeq \Sigma_{\rm mol}$), and low pressure
  ($\Sigma_{\rm gas} \simeq \Sigma_{\rm atom}$) regimes. }
\label{fig:sklaw}
\end{figure*}

Here we give an overview of the star formation law used in our models.

Following Blitz \& Rosolowsky (2006) we assume that star formation
takes place in dense molecular gas, traced by the Hydrogen Cyanide
molecule (HCN), with a constant star formation efficiency:
\begin{equation}
\label{eq:SKmol}
\frac{\Sigma_{\rm SFR}}{[\rm \Msun \,  pc^{-2}\, Gyr ^{-1}]} = \frac{\tilde\epsilon_{\rm SF}}{[\rm Gyr^{-1}]} \frac{\Sigma_{\rm mol, HCN}}{[\rm \Msun \, pc^{-2}]},
\end{equation}
where $\tilde\epsilon_{\rm SF} \simeq  10-13 \rm \,{Gyr}^{-1}$ (Gao \& Solomon
2004, Wu \etal 2005).  Expressing  this equation in terms of the total
gas content:
\begin{equation}
\frac{\Sigma_{\rm SFR}}{[\rm \Msun \,  pc^{-2} \, Gyr ^{-1}]} = \frac{\tilde\epsilon_{\rm SF}}{[\rm Gyr^{-1}]} \frac{\Sigma_{\rm gas}}{[\rm \Msun \, pc^{-2}]} \,f_{\rm mol}\, {\cal R}_{\rm HCN},
\end{equation}
where ${\cal R}_{\rm HCN} = \Sigma_{\rm mol, HCN}/\Sigma_{\rm mol}$ is
the ratio between  the dense molecular gas (as traced  by HCN) and the
total molecular gas, and $f_{\rm mol}$ is the molecular gas fraction.

The fraction of  gas that is molecular, $f_{\rm  mol}$, can be defined
in terms  of the  mass ratio between  molecular and atomic  gas, $\cal
R_{\rm mol}$ by
\begin{equation}
\label{eq:fmol}
f_{\rm mol} = \frac{\cal R_{\rm mol}}{{\cal R}_{\rm mol}+1}.
\end{equation}
Empirically Wong \& Blitz (2002) and Blitz \& Rosolowsky (2004; 2006)
have argued that ${\cal R}_{\rm mol}$ is determined to first order by
the mid plane pressure, $P_{\rm ext}$.  We adopt the relation from
Blitz \& Rosolowsky (2006):
\begin{equation}
\label{eq:Rmol}
{\cal R}_{\rm mol} = \frac{\Sigma_{\rm mol}}{\Sigma_{\rm atom}} 
= \left[ \frac{P_{\rm ext}/k}{4.3\pm0.6 \times 10^4} \right]^{0.92\pm0.1},
\end{equation}
where $k$ is Boltzmann's constant, and $P_{\rm ext}/k$ is in cgs units.
For a gas plus stellar disk the mid plane pressure is given, to
within 10\%, by (Elmegreen 1993)
\begin{equation}
\label{eq:Pext}
P_{\rm ext} \simeq \frac{\pi}{2} G \Sigma_{\rm gas} \left [
  \Sigma_{\rm gas} + 
\left(\frac{\sigma_{\rm gas}}{\sigma_{\rm star}}\right) \Sigma_{\rm
  star} \right],
\end{equation}
where $\sigma_{\rm gas}$ and $\sigma_{\rm star}$ are the velocity
dispersions of the gas and stellar disk, respectively. For simplicity,
we will assume $\sigma_{\rm gas}/\sigma_{\rm star}=0.1$.

Based on the arguments and references in Blitz \& Rosolowsky (2006) we
adopt the following fitting function for ${\cal R}_{\rm HCN}$
\begin{equation}
{\cal R}_{\rm HCN} = 0.1 \left( 1 + \frac{\Sigma_{\rm mol}}{[200 \, \rm \Msun \, pc^{-2}]} \right)^{0.4}. 
\end{equation}

In the low pressure regime, $f_{\rm mol} \simeq {\cal R}_{\rm mol}
\propto \Sigma_{\rm gas}^{1.84\pm0.2}$, $\Sigma_{\rm gas} \simeq
\Sigma_{\rm atom}$, and ${\cal R}_{\rm HCN}\simeq 0.1$, and thus in
terms of total gas or atomic gas Eq.~(\ref{eq:SKmol}) asymptotes to
\begin{eqnarray}
\label{eq:SKlow}
\frac{\Sigma_{\rm SFR}}{[\rm \Msun \, pc^{-2} \, Gyr^{-1}]} = 
\frac{\tilde\epsilon_{\rm SF}}{[\rm Gyr ^{-1}]}\frac{1.37\times 10^{-4}}{[\rm \Msun\, pc^{-2}]}\, \nonumber\\
\times \left(\frac{\Sigma_{\rm gas}}{[1 \rm \Msun \, pc^{-2}]}\right)^{2.84}. 
\end{eqnarray}
Or in terms of the molecular gas, Eq.~(\ref{eq:SKmol}) asymptotes to
\begin{equation}
\label{eq:SKlowh2}
\frac{\Sigma_{\rm SFR}}{[\rm \Msun \, pc^{-2} \, Gyr^{-1}]} = 
\frac{\tilde\epsilon_{\rm SF}}{[\rm Gyr ^{-1}]}\, 
\frac{0.1}{[\rm \Msun \, pc^{-2}]}
\frac{\Sigma_{\rm mol}}{[\rm \Msun \, pc^{-2}]}.
\end{equation}

In the high pressure regime, $f_{\rm mol} \simeq 1$, $\Sigma_{\rm
  gas} \simeq \Sigma_{\rm mol}$, and ${\cal R}_{\rm HCN} \propto
\Sigma_{\rm mol}^{0.4}$, and eq.(\ref{eq:SKmol}) asymptotes to the
familiar SK relation
\begin{eqnarray}
\label{eq:SKhigh}
\frac{\Sigma_{\rm SFR}}{[\rm \Msun \,pc^{-2}\, Gyr^{-1}]}  
= \frac{\tilde \epsilon_{\rm SF}}{[\rm Gyr^{-1}]}\frac{0.0120}{[\rm \Msun\,pc^{-2}]}\, \nonumber \\
\times \left(\frac{\Sigma_{\rm gas}}{[1\, \rm \Msun \,pc^{-2}]}\right)^{1.4}.
\end{eqnarray}
Furthermore, with $\tilde\epsilon_{\rm SF}=13 \,\rm Gyr^{-1}$, we
recover the coefficient of $\epsilon_{\rm SF}=0.16 \rm
\,\Msun\,pc^{-2}\,Gyr^{-1}$ of the standard SK relation\footnote{Note
that the coefficient of $\epsilon_{\rm SF} = 0.25 \rm
\,\Msun\,pc^{-2}\,Gyr^{-1}$ from Kennicutt (1998) does not account for
Helium in the gas masses. Correcting for Helium with a factor of 1.36
results in $\epsilon_{\rm SF}=0.16 \rm \,\Msun\,pc^{-2}\,Gyr^{-1}$.}

We implement the star formation recipe given by Eq.(\ref{eq:SKmol})
as follows.  At each time step and annulus in the disk, we calculate
the star formation rate.  Then we use the following approximation
(valid for times steps small compared to the star formation time
scale) to calculate the mass of newly formed stars
\begin{equation}
\Delta M_{\rm star}(R) = A(R) \,\Sigma_{\rm SFR}(R,t)\,\Delta t,
\end{equation}
with $A$ the area of the annulus, and $\Delta t$ the time step
interval. Note that $\Delta t$ is not a constant in our model, it
varies from $\simeq 4 \rm \,Myr$ at high redshifts to $\simeq 100 \rm
\,Myr$ at low redshifts (green dotted line in Fig.~\ref{fig:tcool}).

{Fig.~\ref{fig:sklaw}} shows the radially resolved star formation law
from 100 randomly selected galaxies from our model with energy driven
outflows (EFB) at redshifts $z=0,2,\& 4$. Each dot corresponds to the
star formation rate surface density and gas surface density of a
single radial bin in a single galaxy. Each galaxy has 10's of such
bins. The red lines show the asymptotic behaviour expected from our
model in the high and low pressure regimes. In the high pressure
regime $\Sigma_{\rm gas} \simeq \Sigma_{\rm mol}$, and so the star
formation law ($\Sigma_{\rm SFR} \propto \Sigma_{\rm gas}^n$) for the
molecular and total gas have the same slope of $n=1.4$ (dotted
lines). In the low pressure regime the star formation law for the
molecular gas has a slope of $n=1$ (solid lines), while that of the
atomic and total gas have a slope of $n=2.84$ (dashed lines). The
panels on the left show that there is a maximum surface density
$\simeq 15 \rm \Msun\,pc^{-2}$ that atomic gas reaches (see Dutton
(2009) for examples of atomic and molecular gas density profiles from
our models). This causes the relation between $\Sigma_{\rm SFR}$ and
$\Sigma_{\rm atom}$ to steepen around $\Sigma_{\rm atom}\simeq 10 \rm
\Msun\,pc^{-2}$, and for the slope to reverse sign for $\Sigma_{\rm
  SFR} \gta 10 \,\rm \Msun\,pc^{-2}$. This regime corresponds to small
galactic radii, where the atomic gas densities {\it decrease} towards
the center of the galaxy, but the molecular gas densities and star
formation rate densities are increasing. The features of our star
formation laws closely resemble the observed relations between local
star formation rate surface density and atomic, molecular, and total
gas density as measured by Bigiel \etal (2008), as well as the
theoretical model of Krumholz, McKee, \& Tumlinson (2009).

\section{Hot Mode vs. Cold Mode Accretion}
\label{sec:hotcold}

\begin{figure}
\centerline{
\psfig{figure=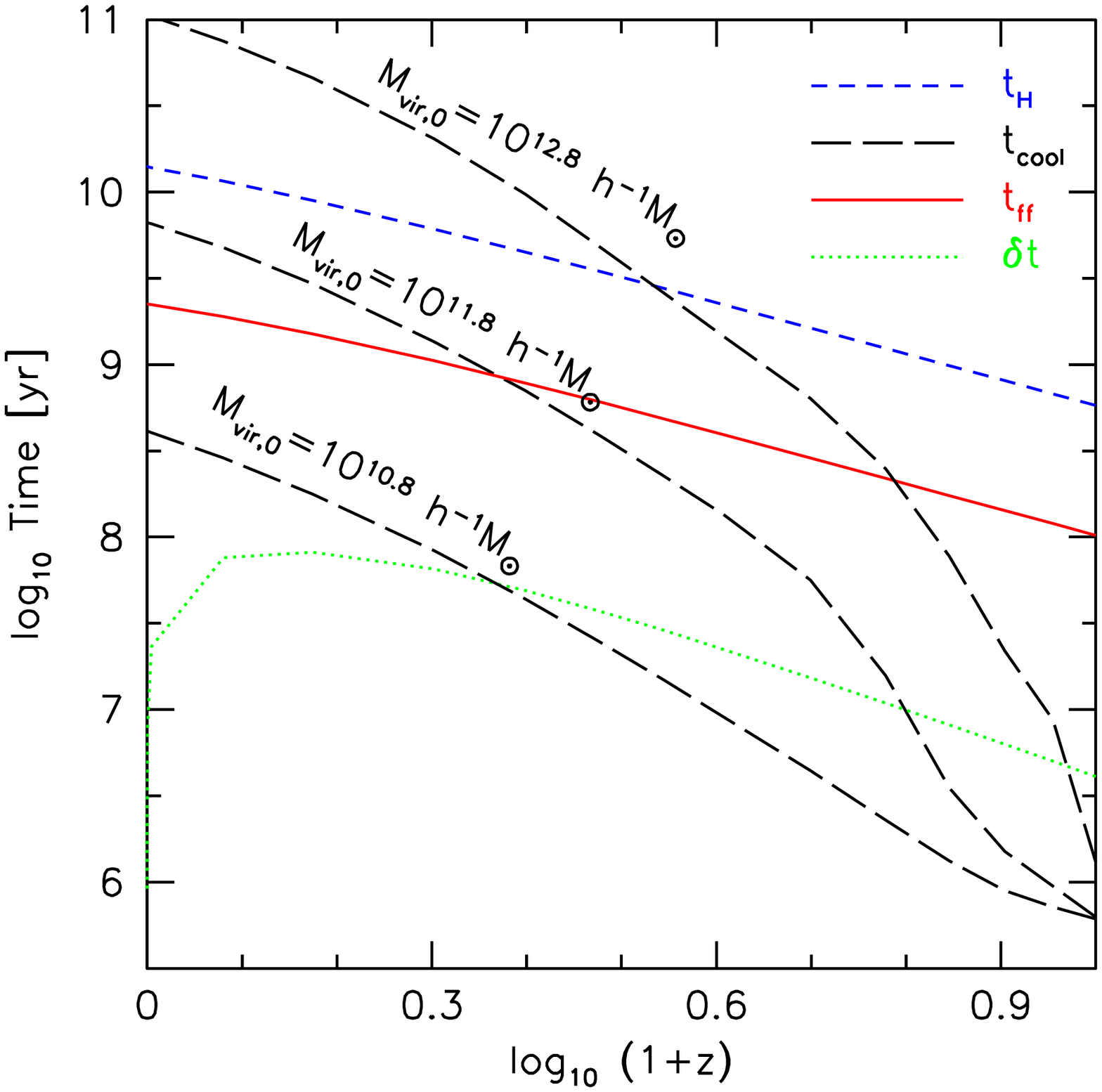,width=0.47\textwidth}}
\caption{Several relevant timescales in our model as function of
  redshift: Hubble time (blue short-dashed); free fall time (red
  solid); cooling times for three halo masses (black long-dashed); and
  numerical time step (green dotted).  Gas accretes onto the central
  galaxy on a timescale that is the maximum of the cooling time and
  the free fall time. At high redshifts the free fall time is the
  limiting timescale, and thus there is a delay of a few 100 Myr
  between when gas enters the halo and when it accretes onto the
  central galaxy. For massive haloes at low redshifts, the cooling
  time exceeds the free-fall time, and may even exceed the Hubble
  time. 
}
\label{fig:tcool}
\end{figure}

It has recently become clear that the way we model the accretion of
gas into galaxies (via a cooling flow of shock heated gas), is not
entirely correct. Birnboim \& Dekel (2003) and Dekel \& Birnboim
  (2006) argued that in haloes below a characteristic mass of $\sim
10^{12} h^{-1} \Msun$, galaxies accrete most of their gas via ``cold
flows'', in which the gas is never heated to the virial temperature of
the halo. This has subsequently been confirmed in cosmological,
hydrodynamical simulations (e.g., Keres \etal 2005).  According to
Dav\'e (2008), the defining features of cold mode accretion are that
it is (i) rapid, (ii) smooth, and (iii) steady.  Rapid because it is
limited by the free fall time, rather than the cooling time, smooth
because most (more than 90\%, Dekel \etal 2009) of the accretion
occurs in small lumps and not through major mergers, and steady
because it is governed by the gravitational potential of the slowly
growing halo.  

What differentiates the cold mode accretion model from the classical
hot mode accretion model? In the hot mode model, gas that enters the
halo is shock heated to the virial temperature. The hot gas then cools
radiatively, and the cold gas accretes onto the central galaxy on a
timescale that is the maximum of the cooling time or the free fall
time.  In cold mode accretion there is either no virial shock, or cold
streams penetrate the hot halo, and gas accretes onto the central
galaxy on a free fall time.

{Fig.~\ref{fig:tcool}} shows the cooling times ($t_{\rm cool}$, black
long-dashed lines) and free fall times ($t_{\rm ff}$, red solid line)
versus redshift for three halo masses ($\Mvir = 10^{10.8,11.8,12.8}
h^{-1} \Msun$) in our adopted cosmology. This shows that for a given
halo, both the cooling time and free fall time decrease to higher
redshift, but that the cooling time decreases faster. Thus at high
redshifts, the free fall time is the limiting factor in the cold gas
accretion rate. For typical halo masses of star forming galaxies
($\Mvir\simeq 10^{11-12}\Msun$) the cooling time is shorter than the
free fall time for most redshifts.

\begin{figure*}
\centerline{
\psfig{figure=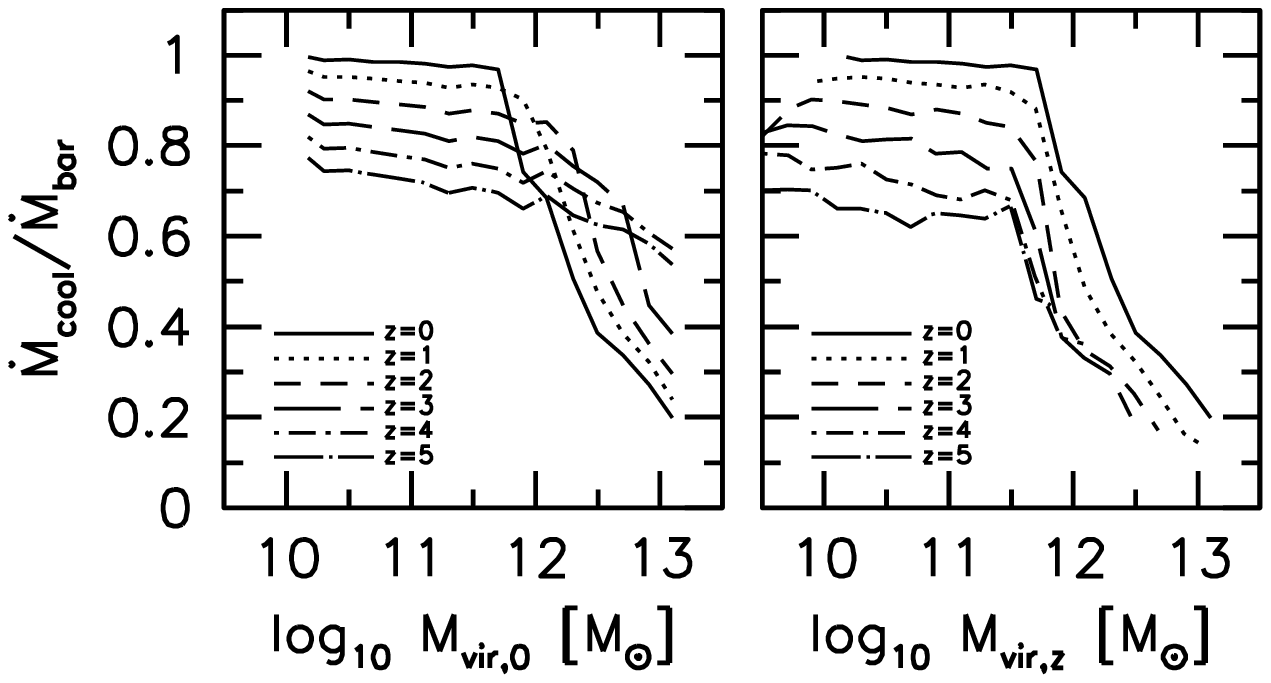,width=0.75\textwidth}}
\caption{Ratio between the accretion rate of cold gas onto the central
  galaxy, $\dot{M}_{\rm cool}$ and the baryon accretion rate into the
  virial radius, $\dot{M}_{\rm bar}$ (the accretion ratio) as a
  function of the virial mass at redshift zero (left-hand panel) and
  as a function of the virial mass at redshift $z$ (right-hand panel).
  The halo mass of $M_{\rm vir,z} \simeq 5\times 10^{11} M_{\odot}$
  marks the transition between fast cooling ($t_{\rm cool} < t_{\rm
    ff}$) and slow cooling ($t_{\rm cool} > t_{\rm ff}$). Hence, our
  hot-mode cooling model yields cold gas accretion rates that are
  indistinguishable from those expected from cold flows.}
\label{fig:mmdot}
\end{figure*}

The impact of this competition between the cooling time and free fall
time on the accretion rates is shown in {Fig.~\ref{fig:mmdot}}.  This
shows the accretion ratio vs. halo mass at redshifts $z=0$ to
$z=5$. The accretion ratio is defined as the ratio between the
accretion rate of cold gas onto the galaxy, $\dot{M}_{\rm cool}$, and
the baryon accretion rate into the halo, $\dot{M}_{\rm bar}$, at that
redshift. As expected from {Fig.~\ref{fig:tcool}}, at low halo masses and
high redshifts haloes are in the fast cooling regime ($t_{\rm cool} <
t_{\rm ff}$), and thus the accretion ratio is close to unity (deviations from
unity arise because of the exponential mass accretion histories adopted here).
Hence, although we do not explicitly model cold accretion flows, effectively
the model yields cold gas accretion rates that are indistinguishable from
those expected from cold flows.

Above a halo mass of $M_{\rm vir,z} \simeq 5\times 10^{11} \Msun$, the
haloes are in the slow cooling regime ($t_{\rm cool} > t_{\rm ff}$),
and the accretion ratio drops sharply. It has been argued that in
haloes above this mass range at high redshifts ($z \gta 2$), the cold
gas accretion rates are dominated by cold streams penetrating the hot
halo (Dekel \etal 2009). This aspect of gas accretion is not captured
by our cooling/accretion treatment, which may impact the formation of
massive galaxies at $z \gta 2$. Apart from this specific case, the
classical hot mode cooling model that we adopt in this paper,
reproduces the defining features of cold mode accretion: rapid,
smooth, and steady (e.g. Dav{\'e} 2008).  We note that while we refer
to cold accretion as smooth, it can also be described as clumpy in the
sense that the accretion is made up of many small lumps, rather than a
diffuse structureless medium.

\section{Why are high redshift disks denser?}
\label{sec:C}
As discussed in \S\ref{sec:dense} the evolution of the densities of
gas disks in our model is stronger than predicted by the simple halo
density scaling.  To understand what causes this deviation we first
discuss what causes the evolution in the surface densities of the
baryonic disks in our model.  Since we assume that the disks are in
centrifugal equilibrium, the evolution in the density of the disk is
determined by the evolution of: 1) the concentration of the halo, $c$;
2) the ratio between the disk mass and total mass, $m_{\rm gal}$; and
3) the ratio between the spin parameter of the disk and the spin
parameter of the halo (i.e. the ratio between the specific angular
momentum of the disk and the specific angular momentum of the halo),
$j_{\rm gal}/m_{\rm gal}$.  Higher $c$, higher $m_{\rm gal}$ and lower
$j_{\rm gal}/m_{\rm gal}$ all result in higher density disks.

To isolate the effects of evolution in $c$ from that of $m_{\rm gal}$
and $j_{\rm gal}/m_{\rm gal}$, we introduce a new model, labeled NFC,
in which we have turned the feedback and cooling calculations off.  In
this model we assume that the ratio between the baryonic mass and
virial mass, $m_{\rm gal}=0.04$, independent of redshift. To implement
this, at each time step, we set the baryonic mass that accretes onto
the galaxy to be $m_{\rm gal}$ times the total mass (baryons plus dark
matter) accreted into the virial radius during that time step. In this
model we also assume that the spin parameter of the halo is
independent of redshift, with a median value of $\lambda=0.035$, and
that the spin parameter of the disk is equal to the spin parameter of
the halo, i.e. $j_{\rm gal}/m_{\rm gal}=1$.

\begin{figure*}
\centerline{
\psfig{figure=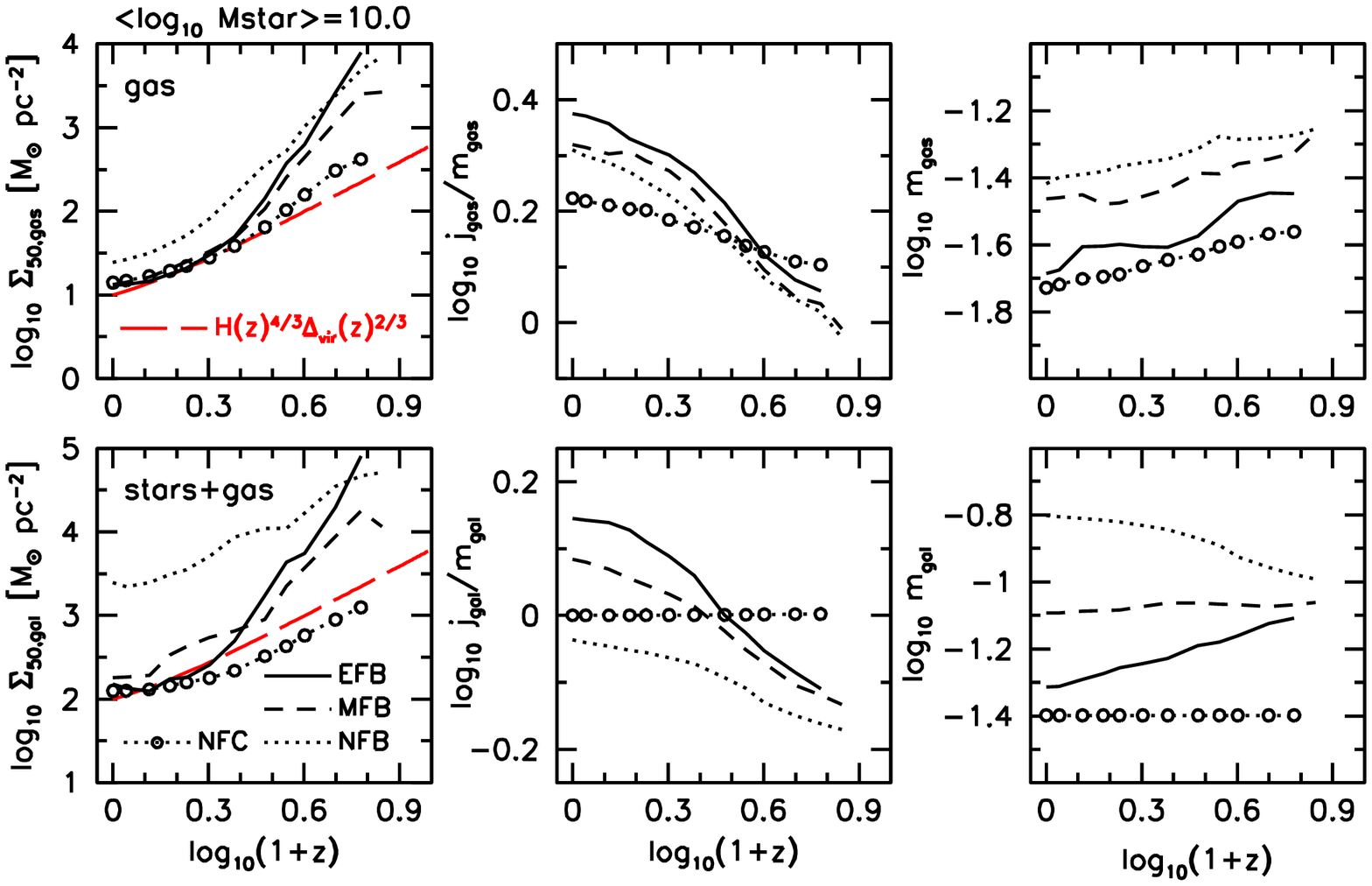,width=0.95\textwidth}}
\caption{ Upper panels: Evolution in the effective surface density of
  the cold gas (left); evolution in the ratio between the specific
  angular momentum of the cold gas to that of the halo (middle);
  evolution in the cold gas mass fraction (right). Lower panels: same
  as upper panels but for cold gas+stars. The deviation from the
  virial scaling (red long dashed lines) is driven by evolution in the
  ratio between the specific angular momentum of the disk and the
  specific angular momentum of the halo. See text for further
  details.}
\label{fig:sgasmz5}
\end{figure*}

The evolution of the surface density of the baryonic disks in the NFC
model is shown as open circles in the lower left panels of
Fig.~\ref{fig:sgasmz5}. This evolution follows the virial scaling for
redshifts $z \gta 1$, but for redshifts $z \lta 1$ the evolution is
weaker. The weak evolution at low redshifts is due to the evolution in
the halo concentration (see Somerville \etal 2008).

The middle and right panels in Fig.~\ref{fig:sgasmz5} show the
evolution of $j_{\rm gal}/m_{\rm gal}$ and $m_{\rm gal}$.  By
construction the NFC model (open circles) shows no evolution in these
parameters. The other three models result in $j_{\rm gal}/m_{\rm gal}$
decreasing with increasing redshift.  For the NFB model the evolution
in $j_{\rm gal}/m_{\rm gal}$ is due to the inside-out nature of the
cooling and accretion of the hot halo gas in our model.  As shown in
Fig.~\ref{fig:tcool}, at high redshifts the cooling time of the hot
halo gas is less than the free fall time, and thus there is a delay
between when baryons enter the halo, and when they reach the central
galaxy. At $z\sim 6$ roughly 60\% of the baryons that accreted into
the halo have cooled and reached the central galaxy.  Since the mass
of the halo is more centrally concentrated than the angular momentum,
by only accreting the inner 60\% of the hot halo baryons, the gas that
reaches the disk has lower specific angular momentum than the halo. By
redshift $z=0$, $m_{\rm gal} \simeq f_{\rm bar}$, and $j_{\rm
  gal}/m_{\rm gal}\simeq 1$.  For the NFB model $j_{\rm gal}/m_{\rm
  gal}$ decreases to higher redshifts, which should result in higher
density disks. However, $m_{\rm gal}$ also decreases to higher
redshifts, which counteracts this effect. This explains why the
evolution of the disk density in the NFB model follows the evolution
of the virial scaling.

At high redshifts all three models with cooling (EFB, MFB, NFB) result
in similar $j_{\rm gal}/m_{\rm gal}$ and $m_{\rm gal}$. This is
because outflows are inefficient in high redshift galaxies due to the
deep potential wells. However, outflows do occur at high redshifts,
which result in the EFB model having lower $m_{\rm gal}$ and higher
$j_{\rm gal}/m_{\rm gal}$ than the MFB and NFB models. Outflows result
in higher $j_{\rm gal}/m_{\rm gal}$ because gas is preferentially
removed from small radii (where the stars are preferentially forming
due to the density dependent star formation law).  At lower redshifts,
the galaxies are less dense, escape velocities are lower, and thus
outflows are more efficient. For the EFB model this results in $m_{\rm
  gal}$ decreasing with decreasing redshift. Thus the combined effects
of lower specific angular momentum and higher mass fractions in higher
redshift disks result in the disk densities increasing faster than the
virial scaling.

Now we return to the evolution of the gas densities. The evolution of
the gas density for the NFC model is shown in the upper left panel in
Fig.~\ref{fig:sgasmz5}. This shows a similar, but slightly stronger
evolution than the virial scaling.  This deviation is qualitatively
explained by the spin parameter of the gas decreasing with increasing
redshift and the gas mass fraction increasing with increasing redshift
(upper middle panel). The evolution of the gas densities in the NFB,
MFB and EFB models is stronger than that of the NFC model because the
specific angular momentum of the disk evolves faster with redshift in
these models.  Thus in our model the gas densities evolve more
strongly than the virial scaling because the spin parameters of the
gas (and baryons) decrease to higher redshifts.

\label{lastpage}

\end{document}